\documentclass[%
 reprint,
 amsmath,amssymb,
 aps,
pra,
floatfix,
]{revtex4-2}

\usepackage{graphicx}
\usepackage{dcolumn}
\usepackage{bm}
\usepackage{braket}
\usepackage{subcaption}
\usepackage{cleveref}
\usepackage{xcolor}

\begin{document}

\title{Enhanced Loading of a Molecular Magneto-Optical Trap}
\author{Ebram Youssef}
\author{Kaiya Wilson}
\author{Jiahe Cao}
\author{Reilly Brislawn}
\author{Avani Lakkireddy}
\author{Kun Liu}
\author{Aaron Teo}
\author{Phoebe Turner}
\author{Loïc Anderegg} \email{anderegg@usc.edu}

\affiliation{Department of Physics and Astronomy, University of Southern California}%

\date{\today}

\begin{abstract}
Molecular magneto-optical traps (MOTs) typically capture orders of magnitude fewer particles than their atomic counterparts due in part to their significantly lower capture velocities. Here, we employ a Stochastic Schrödinger Equation Monte Carlo approach to model a CaF DC MOT to understand the factors limiting capture velocity. We provide physical intuition into the mechanisms that affect capture velocity and identify important parameters and general strategies to improve it. In addition, we point out a loss mechanism intrinsic to molecular MOTs and determine parameter regimes that should be avoided experimentally. We benchmark our simulations against a CaF DC MOT and experimentally implement the improvements predicted by our model. In doing so, we demonstrate a molecular MOT with 1.5 million trapped molecules. This represents an eight-fold improvement and is an important step toward achieving quantum degeneracy with laser cooled molecules.
\end{abstract}

\maketitle

\section{Introduction}
Ultracold molecules have emerged as a versatile system for applications ranging from quantum information processing \cite{DeMille2002,Kaufman2021} and quantum simulation \cite{Micheli2006,Cornish2024}, to ultracold chemistry \cite{Bell2009} and precision tests of fundamental physics \cite{DeMille2024,Hutzler2020}. Driven by these applications, the last decade has seen tremendous progress in the development of laser cooling and trapping techniques for molecules \cite{Barry2014,Truppe2017,Anderegg2017,Collopy2018,Cheuk2018,Caldwell2019,Vilas2022,Burau2023, Zeng2024,PadillaCastillo2025,Hallas2026,Fitch2021,Augenbraun2023,Lasner2025}. However, the complex internal structure of molecules continues to pose challenges, preventing the cooling and trapping of molecules from reaching the temperature and density of their atomic counterparts. The recent development of blue detuned MOTs \cite{Jarvis2018,Burau2023,Hallas2026,Li2025,Zeng2026} has greatly increased molecular MOT densities. However, due to their inherently low capture velocity, blue MOTs must typically be loaded from traditional red MOTs, where the number of particles trapped has remained stagnant \cite{Anderegg2017,Yu2024}. Increasing the number of molecules captured in MOTs enables evaporation required to achieve the long-sought goal of a laser-cooled molecular Bose-Einstein condensate \cite{Bigagli2024,Gadway2016,Langen2025,Schindewolf2025}, provides a direct benefit to the sensitivity of precision searches that employ laser cooled molecules \cite{Safronova2018,Hutzler2020}, and will drastically improve the size of optical tweezer arrays \cite{Anderegg2019,Zhang2022,Holland2023,Vilas2024,Kaufman2021} that can be loaded.

Laser cooled molecules are produced in cryogenic buffer gas beam sources, which generate a high flux molecular beam containing roughly $10^{10}-10^{12}$ \cite{Hutzler2012,Wright2022} molecules per pulse. Of these, roughly $10^{7}-10^{9}$ reach the MOT chamber due to geometric losses, but only $10^4-10^5$ eventually get trapped by the MOT, with the largest molecular MOT to date loading around $2\times10^5$ molecules \cite{Yu2024}. This low yield can be attributed to the inherent inefficiencies of radiative laser slowing and the low capture velocity of molecular MOTs. The laser slowing methods for molecules lack positional dependent slowing, and due to effects such as pluming (loss of low velocity molecules due to the transverse motion), significantly fewer low-velocity molecules arrive into the MOT capture volume compared to their higher velocity counterparts \cite{Barry2012,Hemmerling2016,Truppe2017a}. As a result, a moderate improvement in the capture velocity is expected to lead to a significant improvement in the number of captured molecules.

Due to the complex internal structure of molecules, the changing magnetic and electric fields they experience while moving inside the MOT, the mixing of hyperfine states by these fields, and the many dark states present, modeling the forces and dynamics of a molecular MOT and optimizing its number are not trivial tasks \cite{Tarbutt2015,Tarbutt2015a,Devlin2016,Devlin2018,Langin2023}, neither theoretically nor experimentally. Experimentally, not only is the parameter space vast, but many variables are either inaccessible or technically difficult to tune. Moreover, experimental optimization yields limited insight into the underlying MOT dynamics, often leaving it unclear whether a local or global maximum has been reached.

In this work, we simulate the formation and dynamics of a calcium monofluoride $A^2\Pi_{1/2}$ - $X^2\Sigma^{+}$ DC MOT using the Stochastic Schrödinger Equation (SSE) Monte Carlo approach \cite{Hallas2026,Li2025}. We systematically investigate how the MOT capture velocity and steady-state trapped number depend on a wide range of parameters, such as the intensity, size, and detuning of the various MOT beams and the magnetic field gradient. We characterize the dependence of the capture velocity on experimentally accessible parameters and explain the mechanisms underlying the observed trends. Additionally, we identify a loss mechanism intrinsic to molecular Type-II MOTs and parameters that mitigate it. Finally, we implement the recommended parameters obtained from the simulations in our experimental setup and find good agreement between our simulations and experimental results. This yields a molecular MOT with 1.5 million molecules, an eight-fold improvement in the number of molecules captured over previous MOTs. Although our simulations focus specifically on the internal structure of CaF, underlying principles and optimization strategies are broadly applicable to other laser-cooled molecular species.

\section{Modeling Molecular MOTs}
In order to establish rotationally closed optical cycling for the laser cooling transitions, molecular MOTs must operate between ground states with angular momentum $F$ and an excited state with angular momentum $F'$, where $F'\leq F$. This type of MOT, called a Type-II MOT, introduces dark states to the ground state manifold, which significantly impact the MOT dynamics. Previous theoretical studies have shed light on these MOTs \cite{Devlin2018,Langin2023} and the complex interplay of Sisyphus heating and Doppler cooling. Here, we explore a wider range of parameters relevant to current experimental efforts, such as the intensity of the repump ($v=1$) laser and MOT beam sizes, which we find to play a significant role in the capture velocities of these MOTs. To accurately simulate the dynamics of the CaF MOT, we follow the approaches used by \cite{Li2025,Hallas2026}. 

\subsection{Stochastic Schrödinger Equation}

Laser cooling is a dissipative quantum process in which a local system, the molecule in this case, is coupled to a large reservoir, the quantized EM-vacuum modes in this case, via a dissipative process, which is spontaneous emission. Generally, such open quantum systems can be fully described by their evolution through a master equation, such as the Optical Bloch Equations (OBEs). However, this matrix-based approach can be computationally expensive, as the number of computations in each time step scales as $N^2$ for an N-level system. Another common approach is the rate equations model, where the dynamics of a multi-level system are described by the time evolution of the population in each of the states in the system. Despite its computational efficiency, this approach fails to capture some fundamental effects that can impact the dynamics of laser cooling, such as the formation of dark states from coherent superpositions of states, polarization gradients, and non-steady-state dynamics. A more recently developed method is to numerically solve the stochastic Schrödinger equation, which is also referred to as the Monte Carlo Wavefunction (MCWF) approach \cite{Molmer1993}. This is the method used in this work.

In the MCWF approach, the wavefunction is evolved in time with a non-Hermitian Hamiltonian $H=H_S - \frac{i\hbar}{2}\sum_{m}{C_m^{\dagger}C_m}$, where $H_S$ is the Hamiltonian of the local system and $C_m$ are the collapse operators. This evolution is carried out numerically with small time steps $\delta t$, during which at most one quantum jump event (spontaneous decay in this case) can happen. After every time step, a quantum jump is decided by comparing the probability for a jump with a randomly generated number. The probability of the jump occurring scales with the linewidth of the transition $\Gamma$, as well as the integrated excited state population. When spontaneous decay happens, the direction of the emitted photon, and consequently the momentum kick experienced by the molecules, is determined randomly. The quantum state of the system is then projected to one of the ground states according to the branching ratios, and the polarization of the emitted photon is chosen accordingly. This step simulates the measurement of a photon, in addition to its direction and polarization, and the subsequent collapse of the wavefunction as a result of the measurement. If a quantum jump does not take place, the wavefunction is rotated, effectively increasing the ground state coefficients. The rotated wavefunction is then normalized at the end of each time step. A complete description of the method and its equivalence to master-equation approaches is given in \cite{Dalibard1992, Molmer1993}. Naturally, the computational time in this method scales as $N$ for an N-level system, but due to the stochastic sampling, the simulation might need to be run $n$ times to obtain good statistics of the measured quantities. Depending on the studied system and measured quantity of interest, $n$ might itself be on the order of $N$ or greater. Therefore, the computational efficiency of this method compared to OBEs is situational.

\begin{figure}[t]
    \centering
    \begin{subfigure}{0.475\textwidth}
        \centering
        \includegraphics[width=\linewidth]{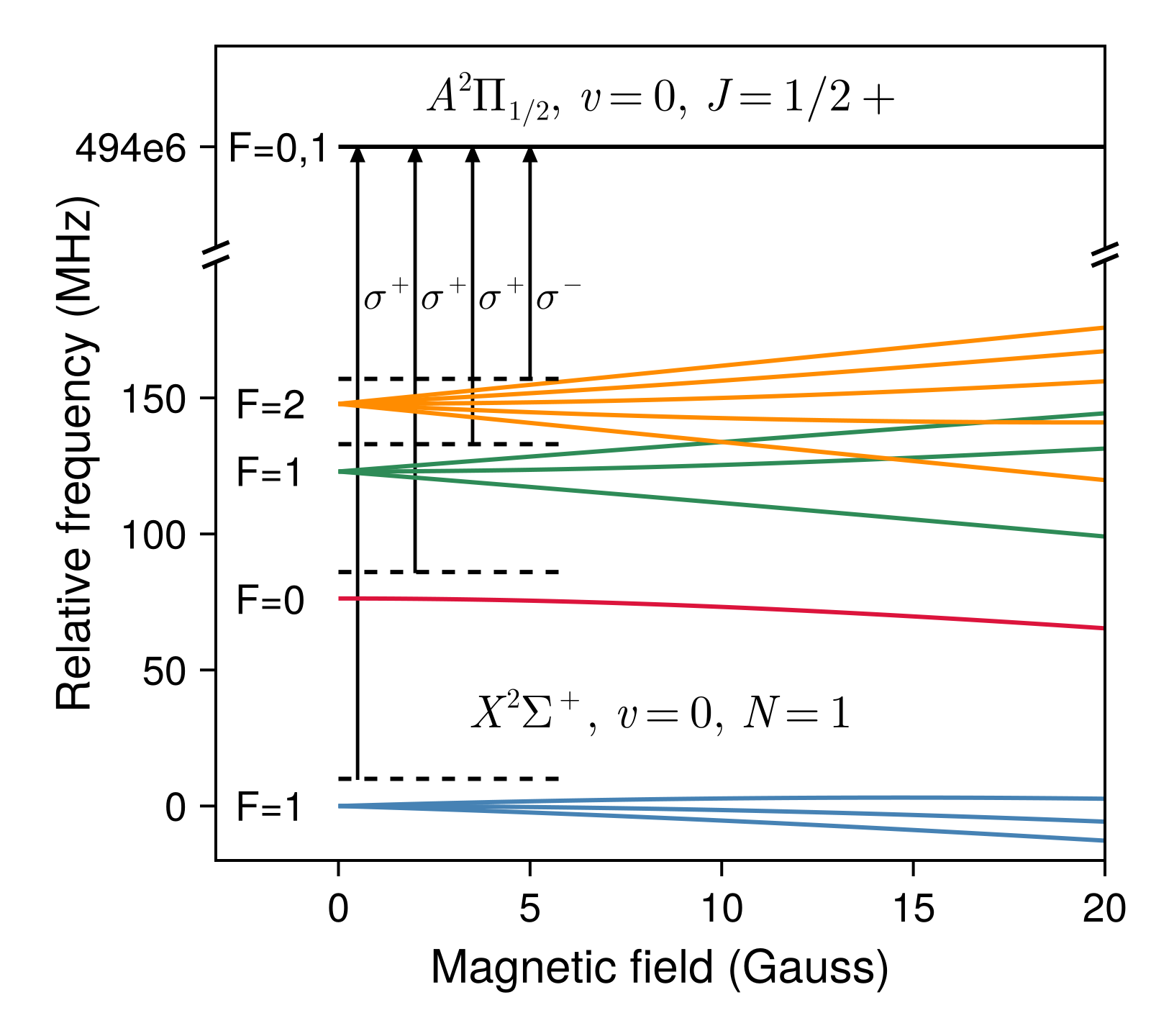}
        \caption{}
        \label{fig:zeemanlevels}
    \end{subfigure}\hfill
    \begin{subfigure}{0.475\textwidth}
        \centering
        \includegraphics[width=\linewidth]{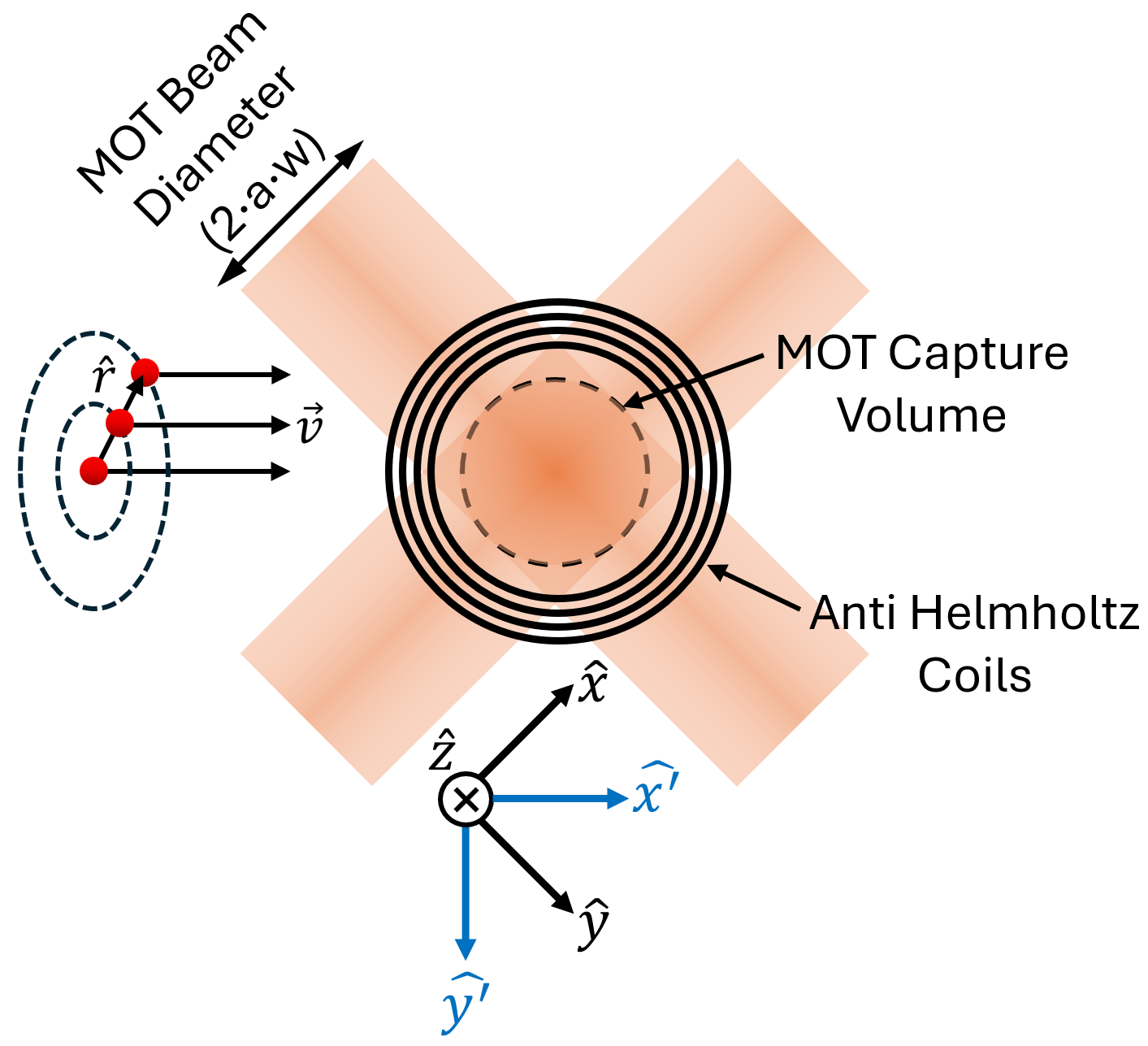}
        \caption{}
        \label{fig:coordinatesystems}
    \end{subfigure}\hfill

    \caption{(a) States used in the simulation and their Zeeman shifts up to 20 Gauss. The v=1 branch of the ground state is not shown but has a hyperfine and Zeeman structure similar to the v=0 branch. Dashed lines represent a detuning of $\Delta=-10$ MHz from the unshifted frequency of each hyperfine level. (b) MOT beam geometry and the coordinate systems referenced in this work.}
    \label{fig:setup}
\end{figure}

\subsection{CaF MOT Simulation Setup}
\label{sec:simulationsetup}

We limit the states used in the simulation to the main laser cooling transition of CaF ($\ket{X^2\Sigma^+, v=0, N=1}$ to $\ket{A^2\Pi_{1/2}, v=0, J=1/2+}$). We also consider the first excited vibrational branch of the ground state $\ket{X^2\Sigma^+, v=1, N=1}$. Each of the two vibrational manifolds contains 12 Zeeman states ($F=1^-, 0, 1^+, 2$ and their $m_F$ levels), while the excited state contains 4 Zeeman states, leading to a system of 28 states. We use the full non-linear Zeeman shifts of the ground states. Due to the small g-factor ($g=-0.02$) of the $A^2\Pi_{1/2}$ state, we leave the excited states unshifted, as such a small shift ($\sim 0.6$ MHz at 20 G) has a negligible contribution to the MOT forces. It is also worth noting that the hyperfine structure of the excited state is unresolved, as the spacing between the two excited hyperfine levels is reported to be $4-5$ MHz, less than the transition linewidth $\Gamma=2\pi~\times8.3$ MHz. Figure \ref{fig:zeemanlevels} shows the ($\ket{X^2\Sigma^+, v=0, N=1}$ to $\ket{A^2\Pi_{1/2}, v=0, J=1/2+}$) energy structure, along with the Zeeman shifts of the v=0 ground states up to 20 Gauss, as well as the laser transitions used in the MOT.

We define a MOT coordinate system $(x, y, z)$, along which the six MOT beams propagate and the magnetic field gradient is defined. The laser beams are defined such that three of them have k-vectors along $+\hat{x}$, $+\hat{y}$, and $+\hat{z}$, then get retro-reflected with opposite circular polarization. The laser beams are assumed to have a Gaussian intensity profile with a $1/e^2$ radius $w$. The intensity profile of each laser beam is set to have a hard cutoff, which mimics the use of an iris, of radius $a\times w$. Unless otherwise noted, $a = 1.5$. This means that the intensity profile from each beam is Gaussian with size $w$ inside $a\times w$ and $0$ otherwise. We use a magnetic field profile $\vec B(x,y,z) = B_{z}(\frac{x}{2} \hat x+\frac{y}{2} \hat y-z \hat z)$, where z is taken to be the axial direction of the anti-Helmholtz coils and $B_z \equiv \partial_{z}B$ is the gradient of the magnetic field along the axial direction. This profile is valid for a field generated by anti-Helmholtz coils near the center of the configuration, and everywhere in space for infinitely large coils. When the full treatment of the B-field profile generated by real anti-Helmholtz coils is used, we see no significant change to the results of these simulations.

In addition, we define a lab coordinate system $(x', y', z') = (\frac{x+y}{\sqrt{2}}, \frac{y-x}{\sqrt{2}}, z)$, which is simply rotated by $45^{\circ}$ from the MOT coordinate system. The molecular beam is set to travel with an initial velocity $v$ towards the center of the MOT along the $+\hat{x'}$ direction. Unless otherwise noted, the initial position of each molecule in the molecular beam is sampled from a uniform distribution of full-widths = (0, 10~mm, 10~mm) about the center $(x'=-3$ cm, $y'=0,~z'=0)$. Figure \ref{fig:coordinatesystems} shows the coordinate systems and MOT beam configuration.

The simulation is run for a time that, based on experimental results and previous simulations, allows enough time for the MOT to form and reach an equilibrium state, which generally is $\sim40~\text{ms}$. At the end of the simulation, the temperature and size of the MOT cloud are calculated by fitting the molecules' final velocities and positions to 3D Maxwell-Boltzmann and Gaussian distributions, respectively. The number of molecules captured and loaded into the steady-state MOT is calculated by counting the molecules that are within the MOT volume (defined by the iris-cutoff size of each MOT beam). When a numerical value for the capture velocity of the MOT is provided, it is calculated by finding the initial beam velocity for which the fraction of captured molecules becomes $0.5$. This is done by fitting a sigmoid function to the captured fraction versus velocity traces. The scattering rate is found by counting the number of collapse events within discrete time bins. In order to benchmark the accuracy of the simulations, we compare the experimentally measured temperature of a red-detuned DC MOT of CaF molecules with the simulation results and find excellent agreement. The results are shown in Appendix \ref{supp:tempcalibration}.

\begin{figure*}[t]

    \centering
    \includegraphics[width=\linewidth]{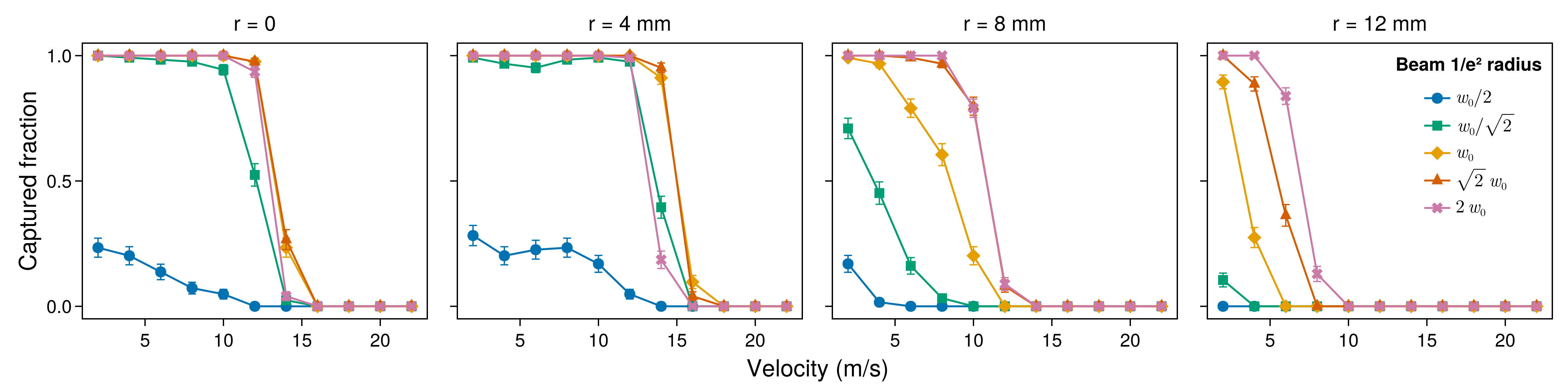}

    \caption{
    \textbf{MOT capture velocity versus beam size at various initial radial displacements.}
    Each subplot displays the fraction of molecules trapped at the simulation's end as a function of laser beam $1/e^2$ radius $w$, radial position $r$, and velocity $v$. Error bars are binomial standard deviations. 
    Fixed parameters: $\Delta = -10~\text{MHz}$, $w_0=8~\text{mm}$, $B_z = -14~\text{G/cm}$, and $P_{v=0} = P_{v=1} = 40~\text{mW}$.}
    \label{fig:radiusvsr0}
\end{figure*}

\section{MOT Loading Optimization}

\begin{figure}
    \centering
    \includegraphics[width=0.9\linewidth]{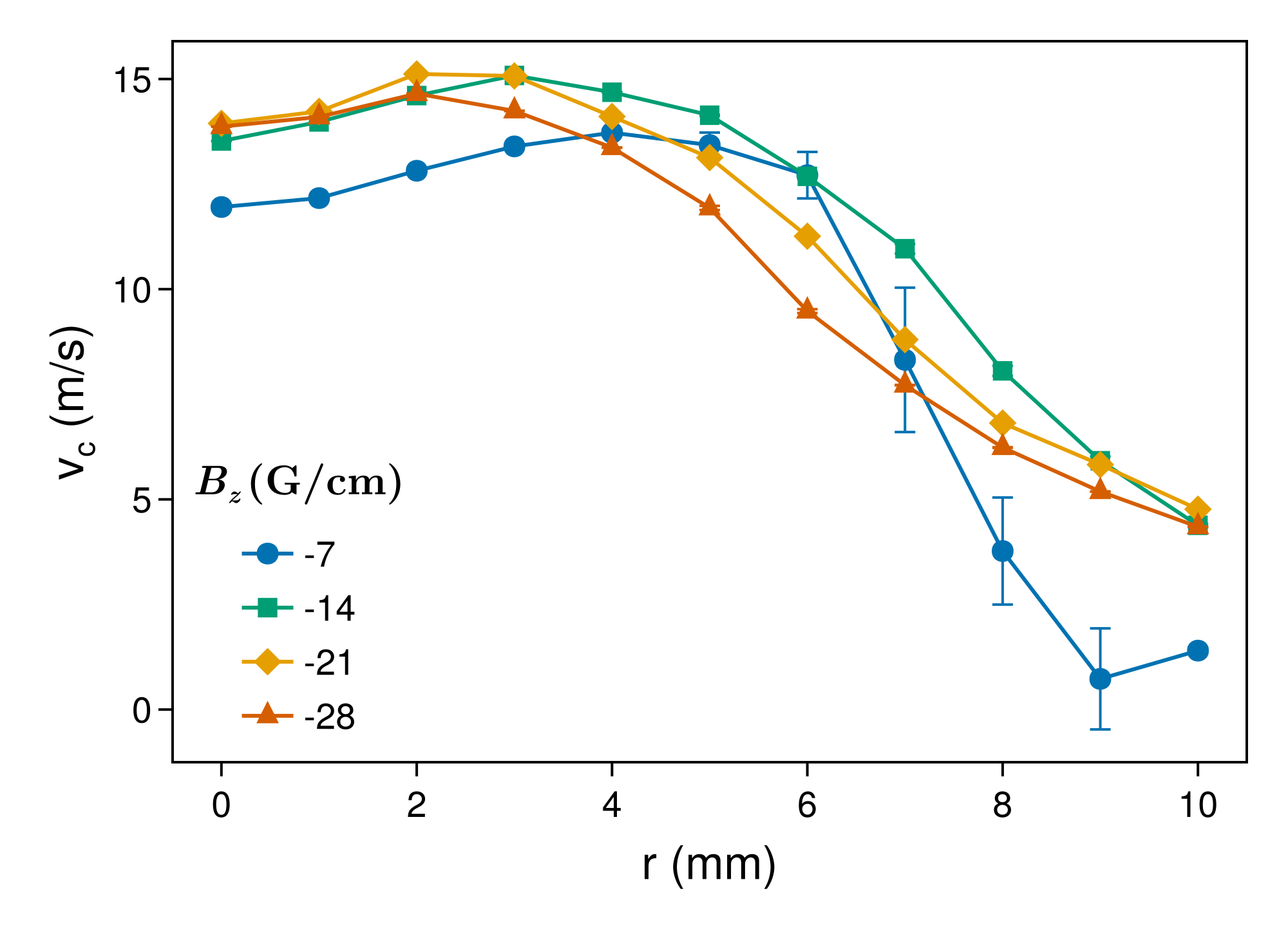}
    \caption{\textbf{MOT capture velocity versus radial displacement at different magnetic field gradients}.
    The $v_c$ values were found by fitting the velocity traces in Fig \ref{fig:radiusvsr0} to a sigmoid function. Error bars are curve fit standard deviations.
    Fixed parameters: $\Delta = -10~\text{MHz}$, $w=8~\text{mm}$, and $P_{v=0}=P_{v=1}=40~\text{mW}$.}
    \label{fig:vc_vs_r0}
\end{figure}

In this section, we scan the main MOT parameters in search of values that maximize the capture velocity and general trends thereof. The main parameters are the laser beam size ($w$), $v=0$ and $v=1$ laser powers ($P_{v=0,1}$), the magnetic field gradient ($B_z$), and the overall detuning of the $v=0$ frequency components from resonance ($\Delta$). For each of the parameter spaces we investigate, we scan the initial velocity $v$ of the molecular beam and record the fraction of molecules that are captured by the MOT, providing a measure of the capture velocity resulting from each set of parameters. Some of the parameters we explore here are easy to optimize experimentally, such as the magnetic field gradient and laser detuning. However, parameters such as the total laser power and the size of the MOT beams are particularly difficult to vary experimentally, making them well suited to be explored in these simulations.

The maximum velocity of a molecule that can be captured into a MOT varies depending on its radial displacement from the $x'$ axis (coinciding with the center of the MOT). In an actual experiment, molecules are loaded from a laser-slowed molecular beam which has a finite spatial extent. The exact size of this slowed molecular beam is experiment dependent. Here, we begin by looking at the dependence of the capture velocity on this radial position.

We consider an ensemble of 124 particles whose initial positions are sampled from a uniform distribution of full-width = 1~mm about $(x_0', y_0', z_0') = (-3~\text{cm}, r/\sqrt{2}, r/\sqrt{2})$, where $r$ is the radial displacement of the molecule from the $x'$ axis. The global v=0 detuning $\Delta$ is set to -10 MHz, which is $\approx 1.2~\Gamma/(2\pi)$. The axial magnetic field gradient $B_z$ is set to -14~G/cm. These values of global detuning and magnetic field gradient are ones that roughly maximize the MOT number in our experiment, making them a good starting point. The total mainline ($v=0$) power per beam $P_{v=0}$ is set to $P_0 = 40~\text{mW}$, which is split among the hyperfine frequency components with ratios equal to the Zeeman degeneracy of each hyperfine level: $(3,1,3,5)/(3+1+3+5)$ for $F=(1^-, 0, 1^+, 2)$, respectively. The hyperfine component polarizations in the z-beam are set to $(\sigma^+, \sigma^+, \sigma^+, \sigma^-)$, which is the configuration that gives rise to the attractive dual-frequency MOT forces \cite{Tarbutt2015a} for the gradient and B-field sign convention we use here. The repump lasers are set to have the same total per-beam power, hyperfine power ratios, and polarizations as the mainline lasers. However, the repump lasers are set to be in resonance with their respective transitions ($\Delta_{v=1}=0$).

\begin{figure*}[tb]
    \includegraphics[width=\linewidth]{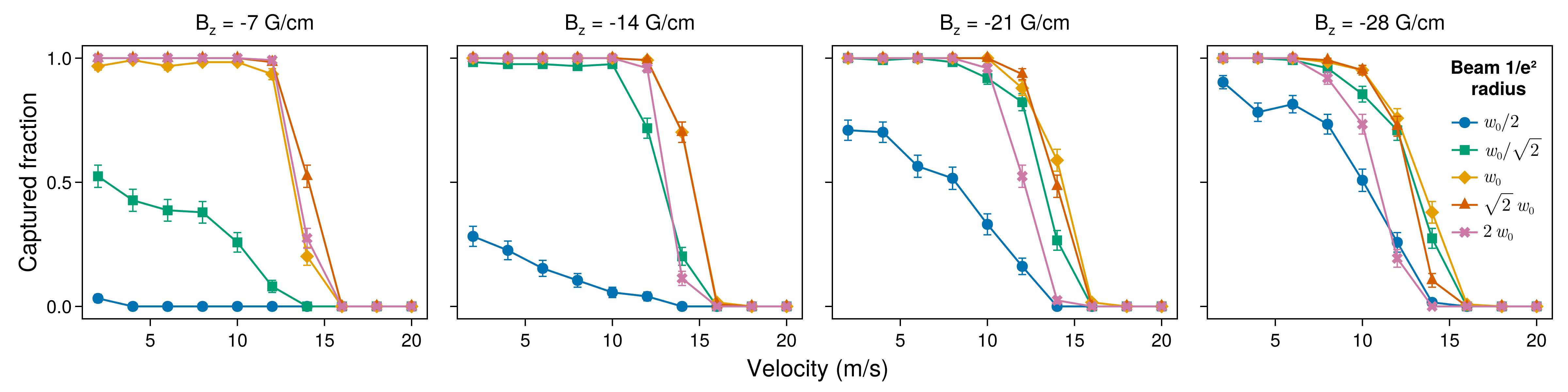}
    \caption{\textbf{MOT capture velocity versus beam size at various magnetic field gradients.} Each subplot displays the fraction of molecules trapped at the simulation's end as a function of laser beam $1/e^2$ radius $w$, magnetic field gradient $B_z$, and velocity $v$. Molecules are uniformly launched over a 1~cm diameter molecular beam. Error bars are binomial standard deviations.
    Fixed parameters: $\Delta=-10~\text{MHz}$, $w_0=8~\text{mm}$, and $P_{v=0}=P_{v=1}=40~\text{mW}$.}
    \label{fig:radiusvsgradient}
\end{figure*}
\begin{figure*}[tb]
    \includegraphics[width=\linewidth]{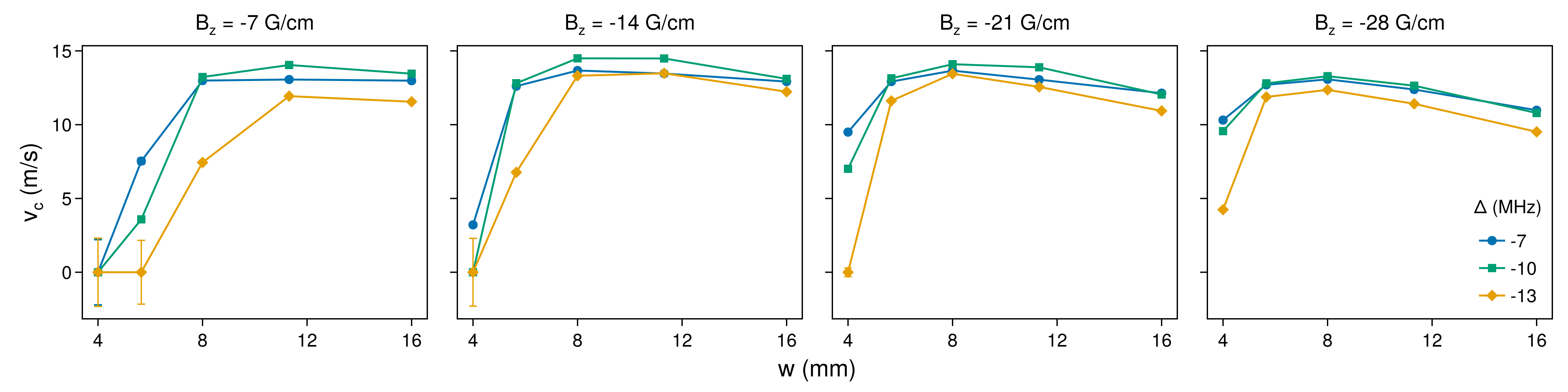}
    \caption{\textbf{MOT capture velocity for various global detunings $\Delta$.} Each subplot displays the capture velocity of the MOT as a function of the magnetic field gradient $B_z$, detuning $\Delta$, and beam radius $w$. The $v_c$ values were found by fitting the velocity traces in Appendix \ref{supp:otherdata} Figure \ref{fig:fullradiusvsgradientvsdetuning} to a sigmoid function. Error bars are curve fitting standard deviations.
    Fixed parameters: $P_{v=0}=P_{v=1}=40~\text{mW}$.}
    
    \label{fig:radiusvsgradientvsdetuning}
\end{figure*}

Figure \ref{fig:radiusvsr0} shows the captured fraction of different initial velocities at various $1/e^2$ beam radii $w$ and initial radial positions $r$. At this set of fixed power, gradient, and detuning, we observe that changing the beam size generally has a small effect on the capture velocity for molecules with a small radial displacement ($r=0-4~\text{mm}$). However, there seems to be a cutoff size below which the capture velocity becomes effectively 0 at all values of $r$. In this case, the size is $w_0/\sqrt2\approx5.7~\text{mm}$. This cutoff effect will be discussed in detail in Section \ref{sec:beamsizethreshold}. 

When $r\sim w$ ($r=8-12~\text{mm}$), the molecules travel through a lower intensity region than those on axis, leading to a global decrease in the capture velocity. Furthermore, for $r=12~\text{mm}$, $y'_0=z'_0\approx8.5~\text{mm}$, which is larger than or equal to the clipped beam size ($a\times w$) for $w=w_0/\sqrt2$ and $w_0/2$, causing the molecules to completely miss the MOT volume. In this large r limit, large MOT beams become clear favorites, even though the average intensity over the MOT is greatly reduced.

We also note that the overall capture efficiency seems to increase slightly for the $r=4~\text{mm}$ molecules compared to on-axis ones. We attribute this to a combination of two effects. First, the radial displacement results in the molecules entering through a region of a larger magnetic field, inside which the Zeeman shifts are more favorable for the dual-frequency force and the overall scattering rate (See Figure \ref{fig:accelerationr0}-\subref{fig:rvsgradientscattering} in Appendix \ref{supp:otherdata}). As a secondary effect, the small $y'$ displacement results in the molecules traveling through a larger cross-section of the $x$ and $y$ MOT beams, which increases the path length along which the high-velocity molecules can be decelerated and turned around (See Figure \ref{fig:intensityvsr} in Appendix \ref{supp:otherdata}). The first effect can be isolated by looking at the capture velocity as a function of $r$ at different gradients, which is shown in Figure \ref{fig:vc_vs_r0}. As the gradient increases, the $v_c$ peak shifts towards a smaller $r$, at which the Zeeman shifts are more favorable for this value of the gradient.

\begin{figure*}[t]
    \includegraphics[width=\linewidth]{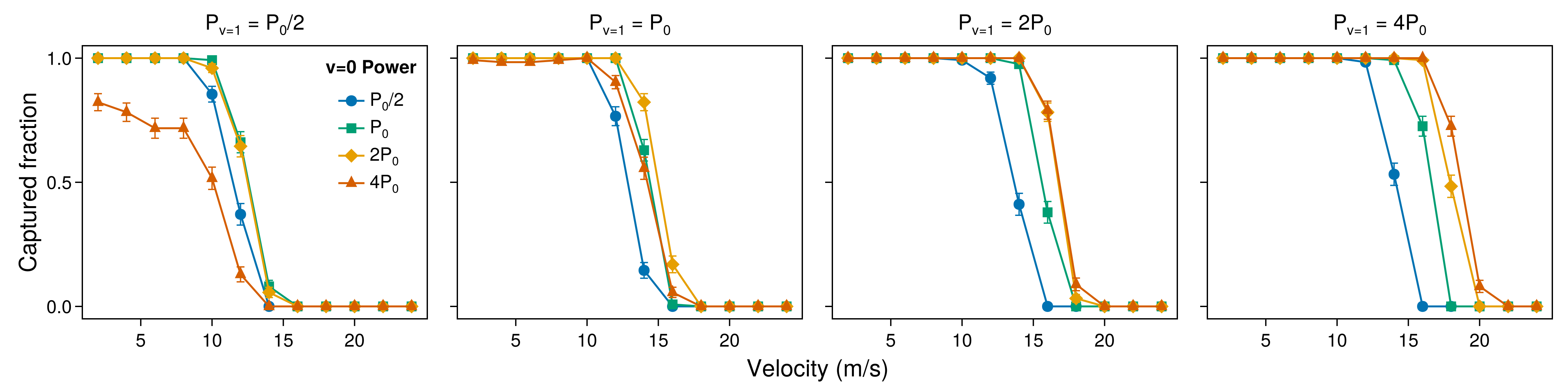}
    \caption{\textbf{MOT capture velocity dependence on mainline power for various repump powers.} Each subplot displays the fraction of molecules trapped at the simulation's end as a function of v=0 power $P_{v=0}$, v=1 power $P_{v=1}$, and molecular beam velocity $v$. Molecules are uniformly launched over a 1~cm diameter molecular beam. Error bars are binomial standard deviations.
    Fixed parameters: $\Delta=-10~\text{MHz}$, $B_z = -14~\text{G/cm}$, $w=8~\text{mm}$, and $P_0=40~\text{mW}$.}
    
    \label{fig:p0vsp1}
\end{figure*}

\begin{figure*}[t]
    \includegraphics[width=\linewidth]{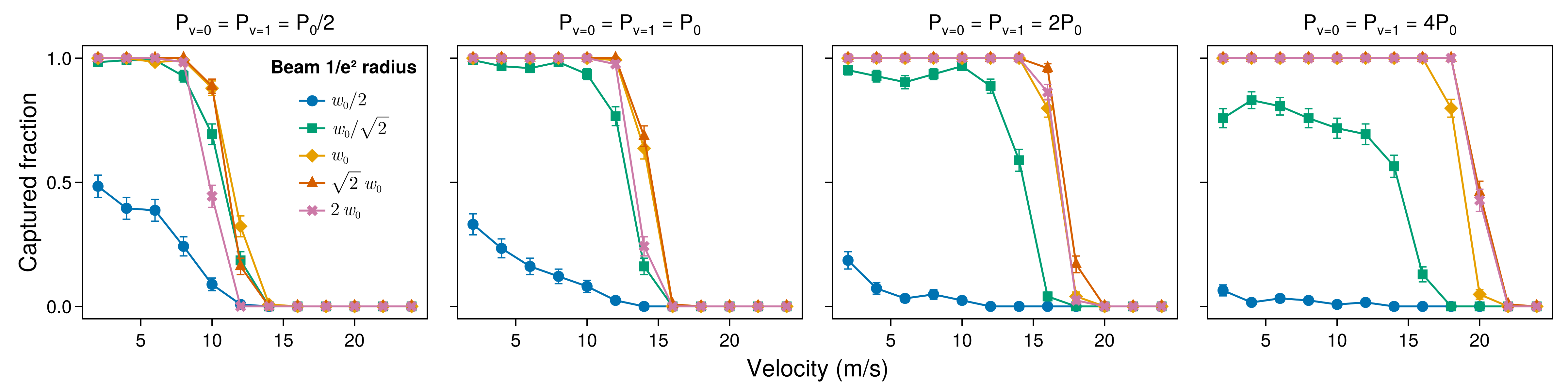}
    \caption{\textbf{MOT capture velocity dependence on beam size for different v=0 and v=1 powers.} Each subplot displays the fraction of molecules trapped at the simulation's end as a function of laser beam size $w$, v=0 and v=1 laser powers (with a 1:1 ratio), and molecular beam velocity $v$. Molecules are uniformly launched over a 1~cm diameter molecular beam. Error bars are binomial standard deviations.
    Fixed parameters: $\Delta=-10~\text{MHz}$, $B_z = -14~\text{G/cm}$, $w_0=8~\text{mm}$, and $P_0=40~\text{mW}$.}
    
    \label{fig:radiusvsp01}
\end{figure*}

In subsequent scans (Figures \ref{fig:radiusvsgradient}-\ref{fig:radiusvsp01}), we will assume a 10~mm diameter slowed molecular beam (representative of a typical experimental slowing beam size) by sampling the initial radial position of the molecules from a 10-mm-wide uniform distribution and plotting the fraction of the molecules captured from the entire diameter. This effectively averages the captured fraction across the entire beam cross-section. While we simulate a molecular beam with a diameter similar to the size of our slowing beam, the precise radial distribution of the slowed molecules is experimentally unknown. Although the optimizations presented below focus on maximizing the capture velocity over this 10~mm range, the MOT loading also depends on the capture area. Larger slowing beams could therefore provide an even more significant advantage, provided a sufficient population of slow molecules is available.

At different MOT beam sizes, we generally expect the molecules to explore different magnitudes of the Zeeman shifts throughout the loading process. For this reason, we next explore the effect of using a different magnetic field gradient on the optimal beam size. In Figure \ref{fig:radiusvsgradient}, we examine the loading efficiency of the MOT as a function of beam size at different magnetic field gradients, while keeping the detuning and powers constant. We find that for the $w_0$ and $\sqrt2w_0$ beams, changing the magnetic field gradient has no significant effect on the MOT capture velocity. Conversely, we see a significant improvement in the capture velocity for the $w_0/\sqrt2$ and $w_0/2$ beams as the gradient is increased. Notably, the $w_0/2$ beam, which effectively does not create a MOT at 7~G/cm, sees a significant increase in capture velocity at the higher gradients, eventually becoming comparable to the larger beams at 28~G/cm. This is related to the confinement of the MOT at a higher gradient and the acceleration experienced as a function of position and velocity, as discussed in Section \ref{sec:beamsizethreshold}. Additionally, the capture velocity for the $2w_0$ beams gradually decreases as the gradient increases. In the presence of a high magnetic field gradient, the ground state Zeeman levels become shifted out of resonance at the outer shells of the MOT, leading to a diminishing force in these regions and limiting the amount of power that contributes to the formation of the MOT (See Figure \ref{fig:largewhighgradientsupport} in Appendix \ref{supp:otherdata}). 

Since the effective shift of the various laser frequencies from resonance depends on both the overall detuning (a constant shift) and the magnetic field gradient (a spatially dependent shift), one might expect the optimal detuning to differ for different magnetic field gradients. In Figure \ref{fig:radiusvsgradientvsdetuning}, we consider the capture velocity of the MOT at different detunings when using the same parameter space we used in Figure \ref{fig:radiusvsgradient}. We see that for beam sizes $\geq w_0$ and gradient magnitudes $\geq 14~\text{G/cm}$, the capture velocity is virtually the same across the considered values of detuning, $\Delta=-7,~-10,~-13~\text{MHz}$. This lack of dependence of the MOT loading on the loading detuning agrees with our experiment for our typical loading parameters ($B_z=-14~\text{G/cm}$, $w=w_0$), where we see approximately the same number of loaded molecules across a loading detuning range of $\sim6~\text{MHz}$. For either a smaller magnitude of the magnetic field gradient ($B_z=-7~\text{G/cm}$) or a smaller beam size ($w\leq w_0/\sqrt2$), we see a greater dependence of the capture velocity on the detuning, where a detuning of either 7 or 10 MHz can be significantly better than 13 MHz. In both cases, the molecules explore a volume of smaller Zeeman shifts in which the transitions are further out of resonance at a larger red detuning.

Thus far, we have used a fixed value for the MOT power that roughly corresponds to the experimentally available amount. We also kept the repump power the same as the mainline power, which is typical of molecular MOT experiments. In Figure \ref{fig:p0vsp1}, we consider the effect of independently changing the mainline and repump powers on the capture velocity when the beam size, gradient, and detuning take on the optimal values found in Figure \ref{fig:radiusvsgradientvsdetuning}. 

\begin{figure*}[t]
    \includegraphics[width=\linewidth]{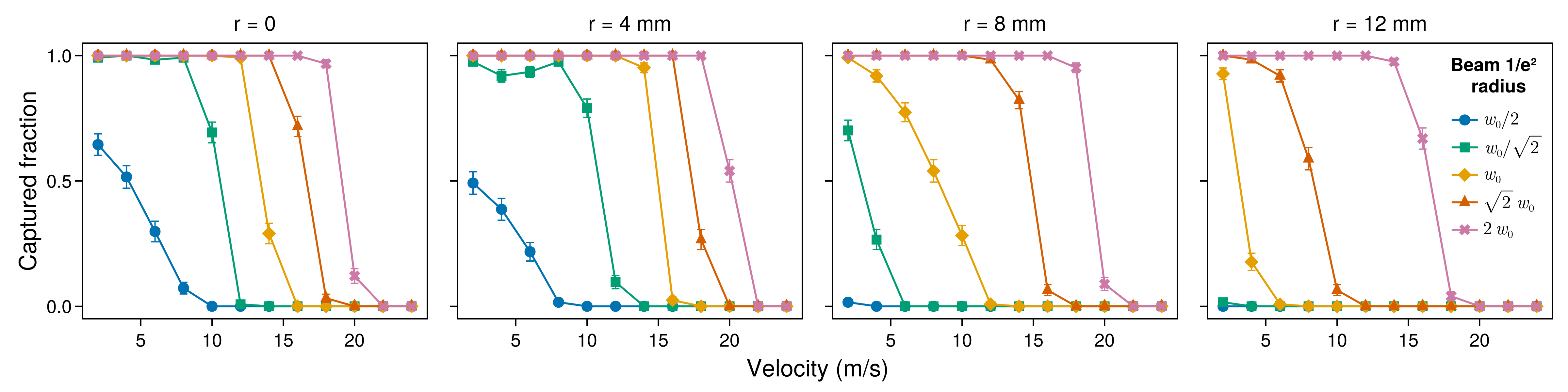}
    \caption{\textbf{MOT capture velocity dependence on beam size at a constant laser intensity for various initial radial displacements.} Each subplot displays the fraction of molecules trapped at the simulation's end as a function of laser beam size $w$, radial displacement $r$, and molecular beam velocity $v$. The peak intensity is kept constant at $I_0$, corresponding to 40~mW over an 8~mm radius laser beam, across the different beam sizes. Error bars are binomial standard deviations.
    Fixed parameters: $\Delta=-10~\text{MHz}$, $B_z = -14~\text{G/cm}$, $w_0=8~\text{mm}$, and $I_0=40~\text{mW/cm$^2$}$.}
    
    \label{fig:radiusvsr0_constintensity}
\end{figure*}

When the repump power is low ($P_{v=1} = P_0/2$), increasing the mainline power from $P_0/2$ to $2P_0$ has little effect on the capture velocity. When the mainline power is further increased to $4P_0$, we see a significant decrease in the capture velocity. While this may appear counterintuitive, we explain the origin of this effect in Section \ref{sec:beamsizethreshold}. When the repump power is increased to $P_0$, the average capture velocity (across the four mainline powers shown) increases, but the decline in capture velocity at $P_{v=0}=4P_0$ is still present. When we increase the repump power to $2P_0$, we see a global increase in the capture velocity across the range of mainline powers, and the decline at $P_{v=0}=4P_0$ vanishes. Eventually, when the repump power is increased to $4P_0$, the $P_{v=0}=4P_0$ capture velocity becomes slightly better than $P_{v=0}=2P_0$.

Interestingly, the $1P_0:4P_0$ mainline to repump ratio sees an increase of 2 m/s in the capture velocity when compared to the $1P_0:1P_0$ configuration, whereas the $4P_0:1P_0$ ratio has no improvement compared to $1P_0:1P_0$. Ultimately, the $4P_0:4P_0$ configuration reaches a capture velocity of 18.6 m/s, 4.4 m/s higher than $1P_0:1P_0$. 

This aligns with previous observations in molecular MOTs, where increasing the mainline power did not lead to an increase in the number of trapped molecules (see Section \ref{sec:experimentalrealization}). This also points to the interesting prospect of increasing the MOT capture velocity solely by increasing the repump power, even when the total experimental mainline power is limited. We expect the noticeable improvement in the MOT capture velocity shown here to lead to a larger number of molecules loaded into the MOT, which we tested experimentally, and the results are discussed in Section \ref{sec:experimentalrealization}.

We then return to the question of what the optimal MOT beam size is given a finite amount of laser powers. In Figure \ref{fig:radiusvsp01}, we consider the capture velocity of the MOT as a function of MOT beam size at different powers of both the $v=0$ and $v=1$ lasers, maintaining a mainline to repump ratio of 1:1. For $P_0/2$ and $P_0$, the $w_0$ and $\sqrt2w_0$ size beams create equally good MOTs, whereas the $2w_0$ size beam, whose intensity is 4 times smaller than the $w_0$ beam, is slightly worse. When the power is increased to $2P_0$, the $2w_0$ size comes to equal footing with the $w_0$ and $\sqrt2w_0$ sizes.

As the power is further increased to $4P_0$, the $2w_0$ and the $\sqrt2w_0$ beams see a noticeable increase in the capture velocity compared to the $w_0$ beam. At $4P_0$, the $2w_0$ and $\sqrt2w_0$ beams have a peak intensity of $I_0$ and $2I_0$, respectively, where $I_0$ is the peak intensity at $P_0$ and $w_0$. Importantly, the $2w_0$ and $\sqrt2w_0$ beams create a MOT volume that is 8 and 3 times the volume created by the $w_0$ beam, respectively. Increasing the MOT volume while maintaining the peak scattering rate becomes advantageous for loading; firstly, it provides a sufficiently large space for the high velocity molecules to be decelerated to equilibrium velocities before reaching the edge of the MOT volume. Additionally, the high power and low magnetic field gradient typically used during loading result in a hot and large MOT cloud. When the MOT volume is sufficiently large compared to the size of this cloud, it minimizes the chance of a trapped molecule escaping the MOT volume through thermal random walk. It is important to note that we used an iris factor of $a=1.5$ for all beam sizes. Experimentally, this is not necessarily the case, since the beam size can become limited by the aperture of the optics (e.g., 1-inch mirrors) as the Gaussian beam size increases. As a result, larger beams experience increased clipping, thereby reducing the available laser power for forming the MOT.

At high power, we see a decline in capture efficiency for the $w_0/\sqrt2$ and $w_0/2$ beams. For the $w_0/2$ beams, this decline is monotonic with increasing power, despite the increased scattering rate. This is due to heating effects that arise at high intensity, as will be discussed in Section \ref{sec:beamsizethreshold}. For the $w_0/\sqrt2$ beams, the capture velocity increases when the powers are increased up to $2P_0$, then declines at $4P_0$ due to the same heating losses. 

The increased MOT volume provided by large MOT beams suggests that the capture cross-section, i.e., the largest radial displacement of an incoming molecule that can still be captured by the MOT, also increases. To investigate this, we perform the same radial displacement scan done in Figure \ref{fig:radiusvsr0}, but keep the intensity constant across the different beam sizes. The intensity used is $I_0=\frac{2P_0}{\pi w_0^2}\approx40~\text{mW}/\text{cm}^2$ per beam per vibrational manifold. This guarantees maintaining the same peak scattering rate and heating effects across all beam sizes. In Figure \ref{fig:radiusvsr0_constintensity}, we see that the larger beams indeed create a larger capture cross-section. For the $2w_0$ beams, the capture velocity remains exceptionally high (16.4 m/s) at $r=12$~mm. If enough molecules enter the MOT at this much radial displacement, the $2w_0$ beams become a clear favorite for loading. Evidently, the optimal MOT beam size depends on the power available, beam cutoff size, and the spatial distribution of the slowed molecular beam near the MOT, all of which are experiment dependent.

We also note that despite the energy crossings between the $F=2$ and $F=1^+$ sublevels starting at 10 Gauss (Figure \ref{fig:zeemanlevels}), the restoring force of the MOT never completely vanishes. Furthermore, an additional dual-frequency force from the $F=1^+$ and the $F=0$ transitions arises at around 30 Gauss, which was pointed out by \cite{Tarbutt2015a}. This allows for the formation of strong confining forces in large MOT beams, given enough power to maintain the peak intensity. In Appendix \ref{supp:otherdata} Figure \ref{fig:2inchbeams}, we find that a capture velocity of 31.5 m/s can be achieved using large MOT beams ($w=22~\text{mm}$) with sufficient powers in a 2-inch optical system. 

\subsection{MOT Beam Size Threshold}
\label{sec:beamsizethreshold}

\begin{figure}[b]
    \centering
    \begin{subfigure}[t]{.68\columnwidth}
        \vspace{0pt}
        \centering
        \includegraphics[width=\linewidth]{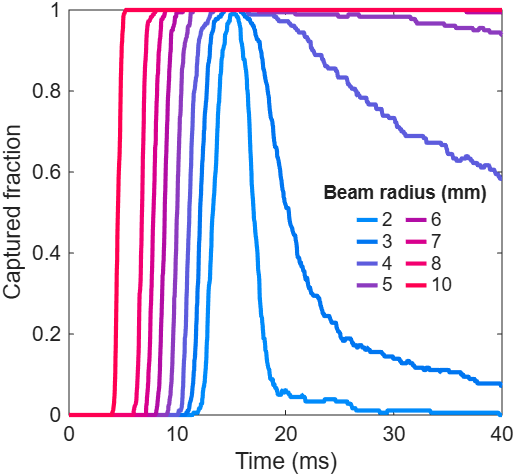}
        \caption{}
        \label{fig:beamsizeloss}
    \end{subfigure}
    \hfill
    \begin{minipage}[t]{.3\columnwidth}
        \vspace{0pt}
        \begin{subfigure}{\linewidth}
            \centering
            \includegraphics[width=\linewidth]{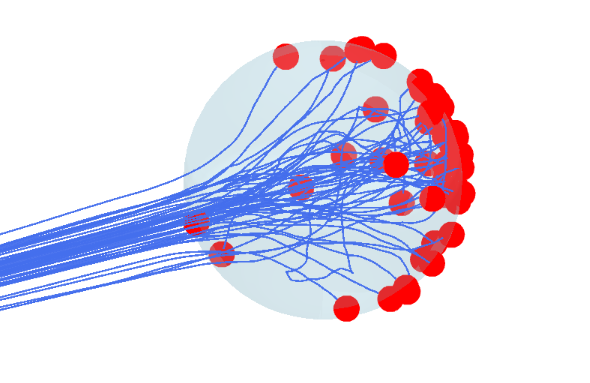}
            \caption{}
            \label{fig:losstraces}
        \end{subfigure}\\
        \vspace{1ex}
        \begin{subfigure}{\linewidth}
            \centering
            \includegraphics[width=\linewidth]{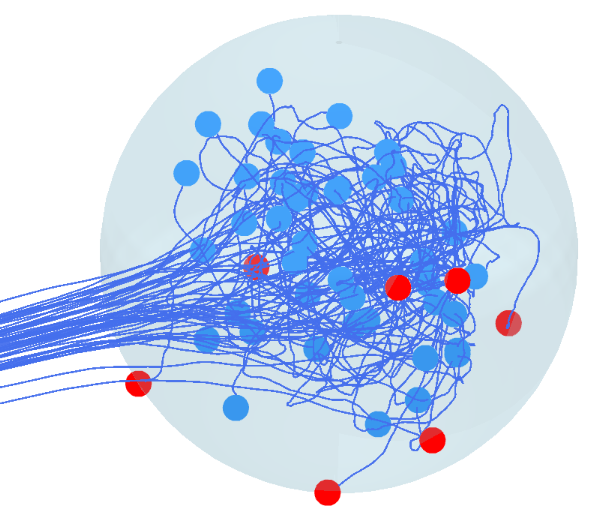}
            \caption{}
            \label{fig:losstraces2}
        \end{subfigure}
    \end{minipage}
      
    \caption{\textbf{Beam size threshold and heating losses.} (a) Time evolution of the captured fraction of 2~m/s molecules as a function of beam size. We use a beam cutoff factor $a=1.5$, which defines the MOT volume as $2\times a\times w$ across each dimension. The number of molecules which are within this volume as a function of time is plotted. (b) Particle trajectories for 3~mm size laser beams and (c) 5~mm size laser beams. Red dots indicate particles that are lost while blue dots are particles that remain in the MOT volume for 40~ms.}
    \label{fig:heatingloss}
\end{figure}

\begin{figure*}[bt]
    \centering
    \includegraphics[width=\linewidth]{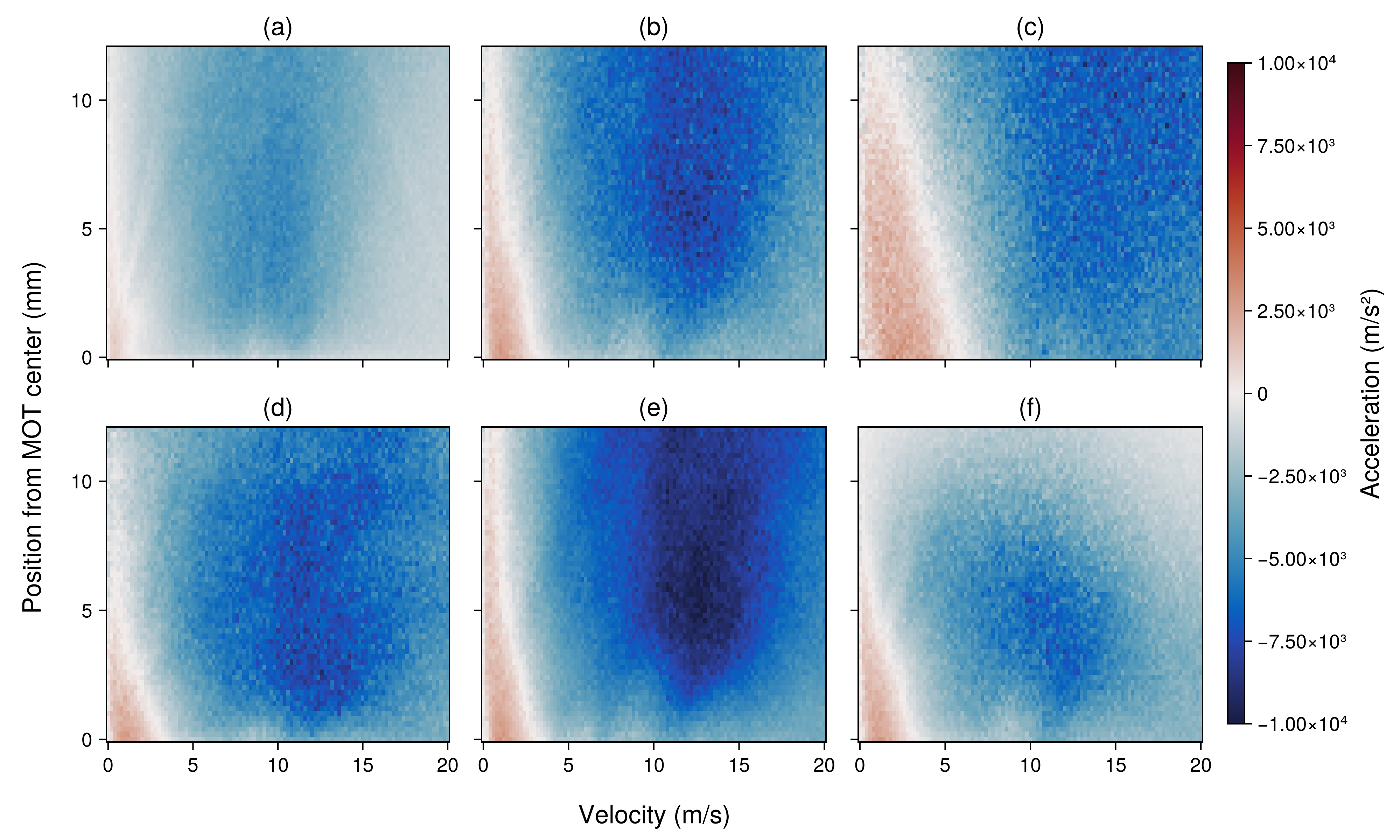}

    \caption{\textbf{Acceleration as a function of molecule velocity and position in the MOT along $\mathbf{\hat{x'}}$.} Infinite beams are used in (a-e) and finite beams are used for (f). Fixed parameters: $I_0=40~\text{mW}/\text{cm}^2$, $\Delta = -10$ MHz, and $B_{z,0}=-14$~G/cm (a) $I_{v=0}=0.25I_0$, $I_{v=1}=I_0$, $B_z=B_{z,0}$ (b) $I_{v=0}=I_{v=1}=I_0$, $B_z=B_{z,0}$ (c) $I_{v=0}=4I_0$, $I_{v=1} = I_0$, $B_z=B_{z,0}$ (d) $I_{v=0}=I_{v=1}=I_0$, $B_z=2B_{z,0}$ (e) $I_{v=0}=I_0$,$I_{v=1}=4I_0$, $B_z=B_{z,0}$ (f) same parameters as (b) but for a finite $w=8$~mm beam.}
    \label{fig:accelerationmaps}
\end{figure*}

In the results obtained thus far, we saw a poor capture efficiency of molecules at all velocities for small MOT beams. This is especially evident at high powers and low magnetic field gradients, where the $w_0/2$ beams fail to capture even the slowest molecules. For example, in Figure \ref{fig:radiusvsr0}, we see a 100\% capture of on-axis molecules up to 10 m/s for a beam size of $w_0/\sqrt{2}$. Based on this, one might expect a molecule moving at 2 m/s (five times slower) to be captured by the $w_0/2$ beams, which are only $\approx30\%$ smaller. However, we see that at this beam size, the capture efficiency is poor at 2~m/s.

To understand this effect, we first examine the time dynamics of the loading process when different MOT beam sizes are used. In the previous results, we only considered if the molecules remained captured after the end of the 40 ms simulation window. If a molecule is not in the MOT volume after 40 ms, it was either never captured or was captured initially and lost later. Figure \ref{fig:beamsizeloss} shows the captured fraction from a molecular beam with velocity $v = 2~\text{m/s}$ as a function of time. One might expect a 100\% capture rate at this velocity for even the smallest MOT beam size plotted ($w=2$~mm), as a CaF molecule only needs to scatter $\sim100$ photons opposing its motion to be slowed down to the Maxwell-Boltzmann velocity corresponding to the equilibrium temperature of a $\sim1~\text{mK}$ red MOT. At a scattering rate of $\sim1$ MHz, the path length at this beam size should be more than sufficiently large to allow for this many photon scatter events. However, for the smallest beams ($w=2-3$~mm), no particles appear to be fully captured, and most molecules fly through the MOT. For slightly larger beams ($w=4-5$~mm), a large portion of the molecules appear captured for a short period, after which there is a slow decay of the captured number. For beam sizes larger than 6~mm, all the molecules remain trapped for the full simulation window.

We can gain a deeper insight by looking at the trajectories of the molecules. In Figure \ref{fig:losstraces}, we see that for the 3~mm MOT beams, all the particles appear to simply enter and exit the MOT without completing a single orbit. By contrast, for the 5~mm beams (Figure \ref{fig:losstraces2}), we see that while a small fraction of the lost molecules are never captured, the majority of the losses occur after the molecules have spent some time in the MOT. The latter loss resembles a boil-off effect, where hot molecules can diffusively escape the rather small MOT volume. This effect can also result in MOT lifetimes that appear shorter than expected based solely on photon budget. 

Although the particle trajectories provide a clearer picture of the nature of the losses present, they do not provide an explanation for the first loss type. That is, why some of these low-velocity molecules never get trapped. For this, we examine the forces experienced by the molecules as a function of position and velocity. Figure \ref{fig:accelerationmaps} shows acceleration maps for various conditions. We see that for low velocities, the MOT is effectively repulsive near the center, and there is a critical position where the force flips sign at each velocity. Thus, if the size of the MOT beams is smaller than the critical position for low-velocity molecules, the molecules are boosted by this anti-restoring force and pushed out of the MOT. The origin of this effect, which was previously explained by \cite{Devlin2018}, is the Sisyphus heating present in red-detuned Type-II MOTs, which is proportional to the AC Stark shifts. This repulsive region increases at higher intensities (Figures \ref{fig:accelerationmaps}a-c). This explains why the cutoff effect becomes more significant as the MOT power is increased (such as in Figure \ref{fig:radiusvsp01}), as shrinking the MOT beams not only reduces the range over which the confining force is present but also increases the intensity.
We also see that by increasing the magnetic field gradient, the extent of this heating feature is reduced (Figure \ref{fig:accelerationmaps}d). This explains the results shown in Figure \ref{fig:radiusvsgradient}, where higher gradients improved the loading for small MOT beams. Finally, we look at the acceleration plot when the repump power is increased four-fold, Figure \ref{fig:accelerationmaps}e. We see that unlike in Figure \ref{fig:accelerationmaps}c, the increase in repump laser power does not create any additional heating, but rather solely increases the restoring force of the MOT. The simulations indicate that the v=0 scattering rate increases with increasing repump power, which might provide an explanation for the boost in restoring force seen here. The v=0 scattering rate as a function of v=0 and v=1 powers is shown in Appendix \ref{supp:otherdata} Figure \ref{fig:simulationscatteringrate}.

While Figures \ref{fig:accelerationmaps}a-e assume infinite-sized laser beams, in actual experiments and all of the previous simulations, the MOT beams are Gaussian with a finite size. Figure \ref{fig:accelerationmaps}f takes into account this finite size. We see that the Gaussian nature of the beam reduces the crossover point where the acceleration switches from anti-restoring to restoring. From the acceleration plot shown for the infinite beam (Figure \ref{fig:accelerationmaps}b), one would expect to need MOT beams larger than 10~mm in radius to create a stable MOT, significantly larger than what is predicted by Figure \ref{fig:losstraces}. This unanticipated advantage of using Gaussian beams should be recognized experimentally, since using a highly clipped Gaussian beam, which creates an effective flat-top profile, would be detrimental to the loading and stability of the MOT.

\begin{figure}[htbp]
    \begin{subfigure}{0.45\textwidth}
        \centering
        \includegraphics[width=\linewidth]{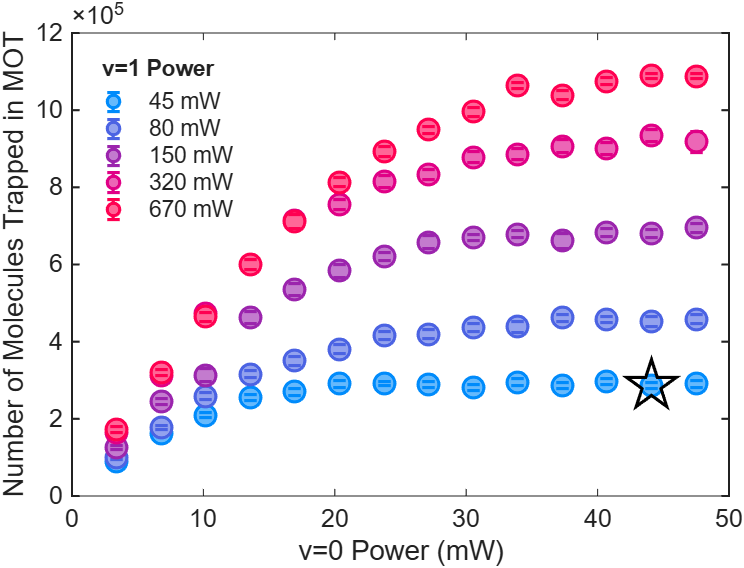}
        \caption{}
        \label{fig:experimentpowerscan}
    \end{subfigure}
    \begin{subfigure}{0.465\textwidth}
        \centering
        \includegraphics[width=\linewidth]{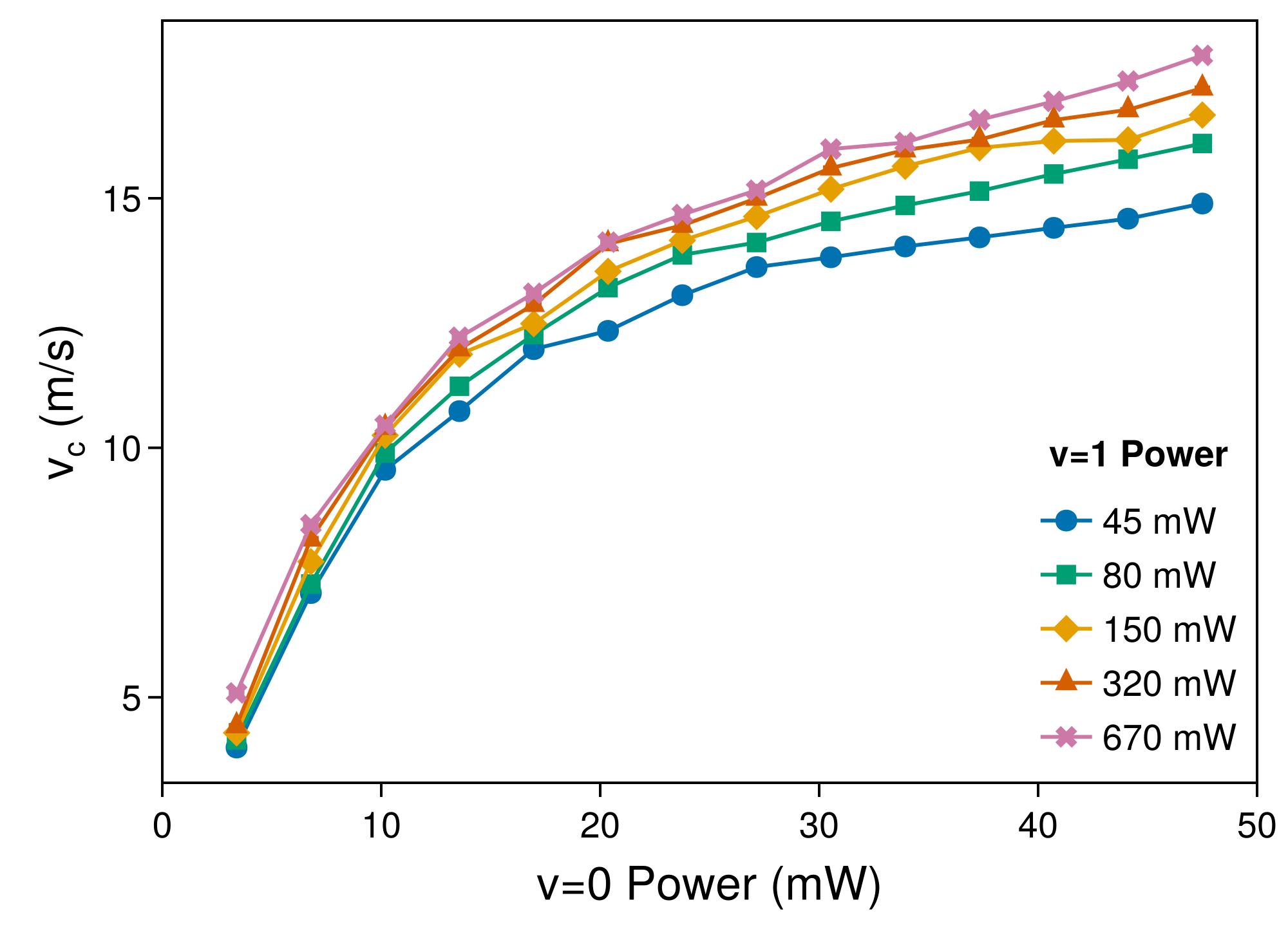}
        \caption{}
        \label{fig:simulationpowerscan}
    \end{subfigure}
    \caption{\textbf{Experimental Realization} (a) \textit{Experimental Data:} Number of molecules loaded into the MOT as a function of X-A v=0 and v=1 power. The pentagram indicates the experimental point nearest to a 1:1 ratio for the v=1 to v=0 power. (b) Simulation results showing the predicted improvement in capture velocity over a 10~mm molecular beam diameter. The $v_c$ values were found by fitting the velocity traces in Appendix \ref{supp:otherdata} Figure \ref{fig:fullP0vsP1expsimdata} to a sigmoid function.}
    \label{fig:powerscancalibration}
\end{figure}

\subsection{Optimization Summary}
To summarize this section, we found that for a fixed amount of laser power, the off-axis capture velocity is significantly increased with a larger MOT beam size, with a small adverse effect to on-axis molecules. Thus, the optimal beam size will be dependent on the size of the slowed molecular beam and the available laser power. When more laser power is available, the larger MOT volume created by larger MOT beams make them have a higher capture velocity regardless of the molecular beam size. The optimal magnetic field gradient and detuning are virtually constant over a wide range of parameters but change noticeably at the beam size extrema. Increasing the mainline laser intensity beyond $80 ~\text{mW/cm}^2$ when the repump laser intensity is limited to less than $20 ~\text{mW/cm}^2$ is detrimental to the MOT capture velocity, as it leads to an increase in the Sisyphus heating over a wide range of positions and velocities. By contrast, increasing the repump power is always beneficial. The Sisyphus heating and anti-restoring forces set a MOT beam size threshold below which the capture velocity drops to zero. This size is determined by the crossover position where the acceleration switches from anti-restoring to restoring.

\section{Experimental Realization}
\label{sec:experimentalrealization}

Following the predicted improvement in capture velocity as a result of increasing the repump laser power, we implemented this in our experiment. Following the typical procedure of creating molecular MOTs \cite{Fitch2021}, we produce CaF molecules in a buffer-gas beam source. This produces a molecular beam which is subsequently slowed using chirped laser slowing on the X-B transition \cite{Truppe2017a}. The slowing beam has a $1/e^2$ diameter of $\approx13$mm at the MOT and is weakly focused to about 10~mm at the cell. The slowing light is combined with v=1 light that is white-light broadened to cover the range of Doppler shifts present. The molecules are then loaded into a red detuned DC X-A MOT, as described in the Section \ref{sec:simulationsetup}. 100~mW of v=2 and 5~mW of v=3 repump light are also present inside the MOT and during slowing.

Previously, the v=1 laser for the MOT was overlapped with the mainline laser in an optical setup composed of AOMs that add frequency shifts corresponding to the ground state hyperfine structure of CaF. To increase the available v=1 repump power in the MOT, we bypass this AOM hyperfine setup and add the hyperfine spectrum using a computer-generated arbitrary phase pattern (similar to Serrodyne modulation) that drives a fiber EOM connected directly to the laser seed \cite{Holland2021}, eliminating losses from the AOMs. We then insert this beam into the MOT in a retro-reflected bow-tie configuration to further increase the amount of available power. This allows us to achieve up to 15 times the maximum intensity of the X-A v=0 MOT beams. Both the mainline MOT beams and the v=1 repump beams have a $1/e^2$ beam radius $w=11$mm and are clipped to just under 0.9 inch in diameter due to the size of the optical elements in the beam path.

To check the dependence of MOT loading on the laser powers, we scanned the power of the mainline MOT beams and v=1 repump during the loading portion of the sequence. Following this, the laser powers were ramped to a constant value, at which a fluorescence image of the MOT was taken. Figure \ref{fig:experimentpowerscan} shows the number of molecules trapped in the MOT as a function of both the mainline and repump powers used during the loading. In typical MOT configurations where the repump power is on the order of the mainline power, we see that increasing the mainline power does not improve the number of captured molecules. In fact, decreasing the mainline intensity by up to a factor of 3 has no impact on the number of captured molecules. When the repump power is increased, we see a clear gain in the number of captured molecules. At a 15:1 ratio of repump to mainline power, we are able to achieve an improvement of about 4 to the number of captured molecules over the typical 1:1 power ratio. Additionally, we see that at higher repump powers, increasing the mainline powers provides a more significant gain.

Figure \ref{fig:simulationpowerscan} shows the simulation result for the average capture velocity over a 1~cm molecular beam using the same set of parameters that were used experimentally. We see that the general trend in the simulated capture velocity agrees with the experimental MOT number trend. However, the experimental gain from increasing the mainline power saturates faster than the gain in capture velocity predicted by the simulation. In addition to the capture velocity of the MOT, the number of loaded molecules in the experiment depends on several correlated factors, such as the capture cross-section of the MOT and the radial and velocity distributions of the slowed molecules. Experimentally, we observe that the relative gain from increasing the repump power 15-fold is sensitive to the ablation laser power, focus, and alignment, indicating that the molecular beam characteristics play a critical role in the final MOT capture efficiency. 

Laser slowing is significantly more effective at producing higher velocity molecules than low velocity ones. This is due to several factors, including the preferential pluming of low forward-velocity molecules due to their relatively high transverse velocity. The transverse heating during the slowing process dictates that the slowed molecules will have a relatively high transverse temperature. Previous studies of laser slowing have shown a decrease of molecules approaching zero velocity \cite{Barry2012,Hemmerling2016,Truppe2017a} compared to molecules at slightly higher velocities, like 20~m/s. As a result, even a slight increase in the capture velocity will lead to a substantial increase in the number of loaded molecules, which we see here.

We find that with these optimized parameters ($P_{v=0}=47~\text{mW}$, $P_{v=1}=670~\text{mW}$, $w=11~\text{mm}$), we are able to load 1.5 million CaF molecules into a DC MOT (see Appendix \ref{supp:numbercalibration}), an 8-fold improvement on previous MOTs \cite{Yu2024}. We also point out that the largest reported molecular MOTs to date are RF MOTs, which were shown to load more molecules than the DC MOT scheme used here \cite{Anderegg2017,Langin2023}. While the RF MOT was not simulated or tested here, we expect these improvements to be additive with the enhanced capture efficiency provided by the RF MOT.

\section{Conclusions}

Here, we have presented a detailed theoretical study for optimizing the loading of CaF molecules into a DC MOT and verified some of the predictions experimentally. Using a stochastic Schr\"odinger equation Monte Carlo model, we explored the various parameters that determine the loading dynamics of a molecular MOT. 

We found that parameters such as the MOT laser powers and beam size play a significant role in the capture velocity of the MOT. We found the optimal MOT beam size to depend on several parameters. Generally, given an abundance of laser power, larger MOT beam diameters are advantageous as they substantially improve the capture of off-axis molecules and provide a larger MOT volume. Interestingly, increasing the power of the main cycling transition does not necessarily increase the capture velocity and can, in fact, reduce it as a result of Sisyphus-like heating forces. On the other hand, increasing the power of the first repump transition reliably enhances the capture velocity. Finally, we identify a parameter regime in which molecules are poorly loaded and rapidly escaped the MOT. This occurs when the MOT beam radius subceeds a threshold at which the net force changes sign and becomes repulsive. The extent of this heating effect in small beams can be reduced by using a larger magnetic field gradient and lower laser intensities. 

Guided by these simulations, we implemented the recommended parameters in our experiment. We see a fourfold improvement in the number of molecules captured by the MOT simply by increasing the power of the first repump laser well past the usual $1:1$ ratio of repump to mainline powers. In combination with the larger MOT beams, which were predicted to increase the capture velocity under our experimental conditions, we are able to load 1.5 million CaF molecules into the MOT, the largest molecular MOT reported to date.

While our model specifically focuses on CaF molecules, we believe the underlying trends and scaling behaviors will be similar to most laser cooled molecular species. A similar numerical approach should be employed by other molecular MOT experiments to optimize the experimental parameters according to the available resources. The improvement in the number of trapped molecules will help achieve quantum degeneracy of laser cooled molecules, as well as load larger optical tweezer arrays and directly improve the sensitivity of precision measurements that use laser cooled molecules. Based on our results, the availability of higher power lasers and the use of even larger MOT beams will further increase the capture velocity of molecular MOTs and should greatly increase the number of trapped molecules. Specifically, large MOT beams in a 2-inch optical system with high enough powers can reach a capture velocity of $>30$~m/s, which could provide another order-of-magnitude increase to the MOT number. The improvements reported here can be combined with other proposed ideas that can improve MOT loading, including transverse cooling and efficient laser slowing methods \cite{Langin2023}, providing a viable pathway towards achieving atomic-like numbers of trapped ultracold molecules.

\begin{acknowledgments}
We thank Christian Hallas and Grace Li for valuable discussions during the development of the SSE simulations. We also thank Yicheng Bao for design improvements of the buffer gas beam source and Scarlett Yu for valuable discussions. Finally, we are grateful to Curt Wittig and the Department of Chemistry at the University of Southern California for providing temporary laboratory space in which this research was conducted. We acknowledge the Center for Advanced Research Computing (CARC) at the University of Southern California for providing computing resources that have contributed to the research results reported within this publication.
\end{acknowledgments}
\bibliography{CaFCaptureVelocity}

@Preamble{
"\providecommand{\noopsort}[1]{}"
# "\providecommand{\singleletter}[1]{#1}%"
}

@Article{Anderegg2017,
  author    = {Anderegg, Loïc and Augenbraun, Benjamin L. and Chae, Eunmi and Hemmerling, Boerge and Hutzler, Nicholas R. and Ravi, Aakash and Collopy, Alejandra and Ye, Jun and Ketterle, Wolfgang and Doyle, John M.},
  journal   = {Physical Review Letters},
  title     = {Radio Frequency Magneto-Optical Trapping of CaF with High Density},
  year      = {2017},
  issn      = {1079-7114},
  month     = sep,
  number    = {10},
  pages     = {103201},
  volume    = {119},
  doi       = {10.1103/physrevlett.119.103201},
  publisher = {American Physical Society (APS)},
}

@Article{Anderegg2019,
  author    = {Anderegg, Loïc and Cheuk, Lawrence W. and Bao, Yicheng and Burchesky, Sean and Ketterle, Wolfgang and Ni, Kang-Kuen and Doyle, John M.},
  journal   = {Science},
  title     = {An optical tweezer array of ultracold molecules},
  year      = {2019},
  issn      = {1095-9203},
  month     = sep,
  number    = {6458},
  pages     = {1156--1158},
  volume    = {365},
  doi       = {10.1126/science.aax1265},
  publisher = {American Association for the Advancement of Science (AAAS)},
}

@Article{Barry2014,
  author    = {Barry, J. F. and McCarron, D. J. and Norrgard, E. B. and Steinecker, M. H. and DeMille, D.},
  journal   = {Nature},
  title     = {Magneto-optical trapping of a diatomic molecule},
  year      = {2014},
  issn      = {1476-4687},
  month     = aug,
  number    = {7514},
  pages     = {286--289},
  volume    = {512},
  doi       = {10.1038/nature13634},
  publisher = {Springer Science and Business Media LLC},
}

@Article{Collopy2018,
  author    = {Collopy, Alejandra L. and Ding, Shiqian and Wu, Yewei and Finneran, Ian A. and Anderegg, Loïc and Augenbraun, Benjamin L. and Doyle, John M. and Ye, Jun},
  journal   = {Physical Review Letters},
  title     = {3D Magneto-Optical Trap of Yttrium Monoxide},
  year      = {2018},
  issn      = {1079-7114},
  month     = nov,
  number    = {21},
  pages     = {213201},
  volume    = {121},
  doi       = {10.1103/physrevlett.121.213201},
  publisher = {American Physical Society (APS)},
}

@Article{Cornish2024,
  author    = {Cornish, Simon L and Tarbutt, Michael R and Hazzard, Kaden RA},
  journal   = {Nature Physics},
  title     = {Quantum simulation with ultracold molecules},
  year      = {2024},
  number    = {5},
  pages     = {730--740},
  volume    = {20},
  doi       = {10.1038/s41567-024-02448-2},
  publisher = {Nature Publishing Group},
}

@Article{DeMille2024,
  author    = {DeMille, David and Hutzler, Nicholas R and Kozyryev, Ivan and Ospelkaus, Christian},
  journal   = {Nature Physics},
  title     = {Quantum precision measurement with molecules},
  year      = {2024},
  number    = {5},
  pages     = {741--749},
  volume    = {20},
  doi       = {10.1038/s41567-024-02449-1},
  publisher = {Nature Publishing Group},
}

@Article{Hemmerling2016,
  author    = {Hemmerling, Boerge and Chae, Eunmi and Ravi, Aakash and Anderegg, Loic and Drayna, Garrett K and Hutzler, Nicholas R and Collopy, Alejandra L and Ye, Jun and Ketterle, Wolfgang and Doyle, John M},
  journal   = {Journal of Physics B: Atomic, Molecular and Optical Physics},
  title     = {Laser slowing of CaF molecules to near the capture velocity of a molecular MOT},
  year      = {2016},
  issn      = {1361-6455},
  month     = aug,
  number    = {17},
  pages     = {174001},
  volume    = {49},
  doi       = {10.1088/0953-4075/49/17/174001},
  publisher = {IOP Publishing},
}

@Article{Holland2023,
  author    = {Holland, Connor M. and Lu, Yukai and Cheuk, Lawrence W.},
  journal   = {Science},
  title     = {On-demand entanglement of molecules in a reconfigurable optical tweezer array},
  year      = {2023},
  issn      = {1095-9203},
  month     = dec,
  number    = {6675},
  pages     = {1143--1147},
  volume    = {382},
  doi       = {10.1126/science.adf4272},
  publisher = {American Association for the Advancement of Science (AAAS)},
}

@Article{Hutzler2012,
  author    = {Hutzler, Nicholas R. and Lu, Hsin-I and Doyle, John M.},
  journal   = {Chemical Reviews},
  title     = {The Buffer Gas Beam: An Intense, Cold, and Slow Source for Atoms and Molecules},
  year      = {2012},
  issn      = {1520-6890},
  month     = may,
  number    = {9},
  pages     = {4803--4827},
  volume    = {112},
  doi       = {10.1021/cr200362u},
  publisher = {American Chemical Society (ACS)},
}

@Article{Kaufman2021,
  author    = {Kaufman, Adam M and Ni, Kang-Kuen},
  journal   = {Nature Physics},
  title     = {Quantum science with optical tweezer arrays of ultracold atoms and molecules},
  year      = {2021},
  number    = {12},
  pages     = {1324--1333},
  volume    = {17},
  doi       = {10.1038/s41567-021-01357-2},
  publisher = {Nature Publishing Group},
}

@Article{Langen2025,
  author    = {Langen, Tim and Boronat, Jordi and Sánchez-Baena, Juan and Bombín, Raúl and Karman, Tijs and Mazzanti, Ferran},
  journal   = {Physical Review Letters},
  title     = {Dipolar Droplets of Strongly Interacting Molecules},
  year      = {2025},
  issn      = {1079-7114},
  month     = feb,
  number    = {5},
  pages     = {053001},
  volume    = {134},
  doi       = {10.1103/physrevlett.134.053001},
  publisher = {American Physical Society (APS)},
}

@Article{Truppe2017,
  author    = {Truppe, S. and Williams, H. J. and Hambach, M. and Caldwell, L. and Fitch, N. J. and Hinds, E. A. and Sauer, B. E. and Tarbutt, M. R.},
  journal   = {Nature Physics},
  title     = {Molecules cooled below the Doppler limit},
  year      = {2017},
  issn      = {1745-2481},
  month     = aug,
  number    = {12},
  pages     = {1173--1176},
  volume    = {13},
  doi       = {10.1038/nphys4241},
  publisher = {Springer Science and Business Media LLC},
}

@Article{Truppe2017a,
  author    = {Truppe, S and Williams, H J and Fitch, N J and Hambach, M and Wall, T E and Hinds, E A and Sauer, B E and Tarbutt, M R},
  journal   = {New Journal of Physics},
  title     = {An intense, cold, velocity-controlled molecular beam by frequency-chirped laser slowing},
  year      = {2017},
  issn      = {1367-2630},
  month     = feb,
  number    = {2},
  pages     = {022001},
  volume    = {19},
  doi       = {10.1088/1367-2630/aa5ca2},
  publisher = {IOP Publishing},
}

@Article{Vilas2022,
  author    = {Vilas, Nathaniel B. and Hallas, Christian and Anderegg, Loïc and Robichaud, Paige and Winnicki, Andrew and Mitra, Debayan and Doyle, John M.},
  journal   = {Nature},
  title     = {Magneto-optical trapping and sub-Doppler cooling of a polyatomic molecule},
  year      = {2022},
  issn      = {1476-4687},
  month     = jun,
  number    = {7912},
  pages     = {70--74},
  volume    = {606},
  doi       = {10.1038/s41586-022-04620-5},
  publisher = {Springer Science and Business Media LLC},
}

@Article{Vilas2024,
  author    = {Vilas, Nathaniel B. and Robichaud, Paige and Hallas, Christian and Li, Grace K. and Anderegg, Loïc and Doyle, John M.},
  journal   = {Nature},
  title     = {An optical tweezer array of ultracold polyatomic molecules},
  year      = {2024},
  issn      = {1476-4687},
  month     = apr,
  number    = {8007},
  pages     = {282--286},
  volume    = {628},
  doi       = {10.1038/s41586-024-07199-1},
  publisher = {Springer Science and Business Media LLC},
}

@Article{Yu2024,
  author        = {Yu, SS and others},
  journal       = {arXiv preprint arXiv:2409.15262},
  title         = {A conveyor-belt magneto-optical trap of CaF},
  year          = {2024},
  archiveprefix = {arXiv},
  eprint        = {2409.15262},
}

@Article{Micheli2006,
  author    = {Micheli, A. and Brennen, G. K. and Zoller, P.},
  journal   = {Nature Physics},
  title     = {A toolbox for lattice-spin models with polar molecules},
  year      = {2006},
  issn      = {1745-2481},
  month     = Apr,
  number    = {5},
  pages     = {341--347},
  volume    = {2},
  doi       = {10.1038/nphys287},
  publisher = {Springer Science and Business Media LLC},
}

@Article{Gadway2016,
  author    = {Gadway, Bryce and Yan, Bo},
  journal   = {Journal of Physics B: Atomic, Molecular and Optical Physics},
  title     = {Strongly interacting ultracold polar molecules},
  year      = {2016},
  issn      = {1361-6455},
  month     = Jun,
  number    = {15},
  pages     = {152002},
  volume    = {49},
  doi       = {10.1088/0953-4075/49/15/152002},
  publisher = {IOP Publishing},
}

@Article{Hutzler2020,
  author    = {Hutzler, Nicholas R},
  journal   = {Quantum Science and Technology},
  title     = {Polyatomic molecules as quantum sensors for fundamental physics},
  year      = {2020},
  issn      = {2058-9565},
  month     = Oct,
  number    = {4},
  pages     = {044011},
  volume    = {5},
  doi       = {10.1088/2058-9565/abb9c5},
  publisher = {IOP Publishing},
}

@Article{Hallas2026,
  author    = {Hallas, Christian and Li, Grace K. and Vilas, Nathaniel B. and Robichaud, Paige and Anderegg, Loïc and Doyle, John M.},
  journal   = {Physical Review Letters},
  title     = {High Compression Blue-Detuned Magneto-Optical Trap of Polyatomic Molecules},
  year      = {2026},
  issn      = {1079-7114},
  month     = Apr,
  number    = {13},
  volume    = {136},
  doi       = {10.1103/w9qc-rczf},
  publisher = {American Physical Society (APS)},
}

@Article{Li2025,
  author    = {Li, Grace K and Hallas, Christian and Doyle, John M},
  journal   = {New J. Phys.},
  title     = {Conveyor-belt magneto-optical trapping of molecules},
  year      = {2025},
  issn      = {1367-2630},
  month     = apr,
  number    = {4},
  pages     = {043002},
  volume    = {27},
  doi       = {10.1088/1367-2630/adc032},
  publisher = {IOP Publishing},
}

@Article{DeMille2002,
  author    = {DeMille, D.},
  journal   = {Physical Review Letters},
  title     = {Quantum Computation with Trapped Polar Molecules},
  year      = {2002},
  issn      = {1079-7114},
  month     = Jan,
  number    = {6},
  pages     = {067901},
  volume    = {88},
  doi       = {10.1103/physrevlett.88.067901},
  publisher = {American Physical Society (APS)},
}

@Article{Burau2023,
  author    = {Burau, Justin J. and Aggarwal, Parul and Mehling, Kameron and Ye, Jun},
  journal   = {Physical Review Letters},
  title     = {Blue-Detuned Magneto-optical Trap of Molecules},
  year      = {2023},
  issn      = {1079-7114},
  month     = May,
  number    = {19},
  pages     = {193401},
  volume    = {130},
  doi       = {10.1103/physrevlett.130.193401},
  publisher = {American Physical Society (APS)},
}

@Article{Cheuk2018,
  author    = {Cheuk, Lawrence W. and Anderegg, Loïc and Augenbraun, Benjamin L. and Bao, Yicheng and Burchesky, Sean and Ketterle, Wolfgang and Doyle, John M.},
  journal   = {Physical Review Letters},
  title     = {$\Lambda$-Enhanced Imaging of Molecules in an Optical Trap},
  year      = {2018},
  issn      = {1079-7114},
  month     = Aug,
  number    = {8},
  pages     = {083201},
  volume    = {121},
  doi       = {10.1103/physrevlett.121.083201},
  publisher = {American Physical Society (APS)},
}

@Article{Caldwell2019,
  author    = {Caldwell, L. and Devlin, J. A. and Williams, H. J. and Fitch, N. J. and Hinds, E. A. and Sauer, B. E. and Tarbutt, M. R.},
  journal   = {Physical Review Letters},
  title     = {Deep Laser Cooling and Efficient Magnetic Compression of Molecules},
  year      = {2019},
  issn      = {1079-7114},
  month     = Jul,
  number    = {3},
  pages     = {033202},
  volume    = {123},
  doi       = {10.1103/physrevlett.123.033202},
  publisher = {American Physical Society (APS)},
}

@Article{Zeng2024,
  author    = {Zeng, Zixuan and Deng, Shuhua and Yang, Shoukang and Yan, Bo},
  journal   = {Physical Review Letters},
  title     = {Three-Dimensional Magneto-Optical Trapping of Barium Monofluoride},
  year      = {2024},
  issn      = {1079-7114},
  month     = Oct,
  number    = {14},
  pages     = {143404},
  volume    = {133},
  doi       = {10.1103/physrevlett.133.143404},
  publisher = {American Physical Society (APS)},
}

@Article{PadillaCastillo2025,
  author    = {Padilla-Castillo, J. E. and Cai, J. and Agarwal, P. and Kukreja, P. and Thomas, R. and Sartakov, B. G. and Truppe, S. and Meijer, G. and Wright, S. C.},
  journal   = {Physical Review Letters},
  title     = {Magneto-Optical Trapping of Aluminum Monofluoride},
  year      = {2025},
  issn      = {1079-7114},
  month     = Dec,
  number    = {24},
  volume    = {135},
  doi       = {10.1103/ksnd-9fyf},
  publisher = {American Physical Society (APS)},
}

@Article{Zhang2022,
  author    = {Zhang, Jessie T and Picard, Lewis R B and Cairncross, William B and Wang, Kenneth and Yu, Yichao and Fang, Fang and Ni, Kang-Kuen},
  journal   = {Quantum Science and Technology},
  title     = {An optical tweezer array of ground-state polar molecules},
  year      = {2022},
  issn      = {2058-9565},
  month     = May,
  number    = {3},
  pages     = {035006},
  volume    = {7},
  doi       = {10.1088/2058-9565/ac676c},
  publisher = {IOP Publishing},
}

@Article{Safronova2018,
  author    = {Safronova, Marianna S and Budker, Dmitry and DeMille, David and Kimball, Derek F Jackson and Derevianko, Andrei and Clark, Charles W},
  journal   = {Reviews of Modern Physics},
  title     = {Search for new physics with atoms and molecules},
  year      = {2018},
  issn      = {1539-0756},
  month     = Jun,
  number    = {2},
  pages     = {025008},
  volume    = {90},
  doi       = {10.1103/RevModPhys.90.025008},
  publisher = {APS},
}

@Article{Barry2012,
  author    = {Barry, J. F. and Shuman, E. S. and Norrgard, E. B. and DeMille, D.},
  journal   = {Physical Review Letters},
  title     = {Laser Radiation Pressure Slowing of a Molecular Beam},
  year      = {2012},
  issn      = {1079-7114},
  month     = Mar,
  number    = {10},
  pages     = {103002},
  volume    = {108},
  doi       = {10.1103/physrevlett.108.103002},
  publisher = {American Physical Society (APS)},
}

@Article{Devlin2016,
  author    = {Devlin, J A and Tarbutt, M R},
  journal   = {New Journal of Physics},
  title     = {Three-dimensional Doppler, polarization-gradient, and magneto-optical forces for atoms and molecules with dark states},
  year      = {2016},
  issn      = {1367-2630},
  month     = Dec,
  number    = {12},
  pages     = {123017},
  volume    = {18},
  doi       = {10.1088/1367-2630/18/12/123017},
  publisher = {IOP Publishing},
}

@Article{Devlin2018,
  author    = {Devlin, J. A. and Tarbutt, M. R.},
  journal   = {Physical Review A},
  title     = {Laser cooling and magneto-optical trapping of molecules analyzed using optical Bloch equations and the Fokker-Planck-Kramers equation},
  year      = {2018},
  issn      = {2469-9934},
  month     = Dec,
  number    = {6},
  pages     = {063415},
  volume    = {98},
  doi       = {10.1103/physreva.98.063415},
  publisher = {American Physical Society (APS)},
}

@Article{Tarbutt2015,
  author    = {Tarbutt, M R},
  journal   = {New Journal of Physics},
  title     = {Magneto-optical trapping forces for atoms and molecules with complex level structures},
  year      = {2015},
  issn      = {1367-2630},
  month     = Jan,
  number    = {1},
  pages     = {015007},
  volume    = {17},
  doi       = {10.1088/1367-2630/17/1/015007},
  publisher = {IOP Publishing},
}

@Article{Tarbutt2015a,
  author    = {Tarbutt, M. R. and Steimle, T. C.},
  journal   = {Physical Review A},
  title     = {Modeling magneto-optical trapping of CaF molecules},
  year      = {2015},
  issn      = {1094-1622},
  month     = Nov,
  number    = {5},
  pages     = {053401},
  volume    = {92},
  doi       = {10.1103/physreva.92.053401},
  publisher = {American Physical Society (APS)},
}

@Article{Langin2023,
  author    = {Langin, T K and DeMille, D},
  journal   = {New Journal of Physics},
  title     = {Toward improved loading, cooling, and trapping of molecules in magneto-optical traps},
  year      = {2023},
  issn      = {1367-2630},
  month     = Apr,
  number    = {4},
  pages     = {043005},
  volume    = {25},
  doi       = {10.1088/1367-2630/acc34d},
  publisher = {IOP Publishing},
}

@Article{Molmer1993,
  author    = {Klaus Molmer, Yvan Castin, Jean Dalibard},
  journal   = {Journal of the Optical Society of America},
  title     = {Monte Carlo wave-function method in quantum optics},
  year      = {1993},
  issn      = {1520-8540},
  month     = Mar,
  number    = {3},
  pages     = {524},
  volume    = {10},
  doi       = {10.1364/josab.10.000524},
  publisher = {Optica Publishing Group},
}

@Article{Dalibard1992,
  author    = {Jean Dalibard, Yvan Castin, Klaus Molmer},
  journal   = {Physical Review Letters},
  title     = {Wave-Function Approach to Dissipative Processes in Quantum Optics},
  year      = {1992},
  issn      = {0031-9007},
  month     = Feb,
  number    = {5},
  pages     = {580--583},
  volume    = {68},
  doi       = {10.1103/physrevlett.68.580},
  publisher = {American Physical Society (APS)},
}

@Article{Bigagli2024,
  author    = {Bigagli, Niccolò and Yuan, Weijun and Zhang, Siwei and Bulatovic, Boris and Karman, Tijs and Stevenson, Ian and Will, Sebastian},
  journal   = {Nature},
  title     = {Observation of Bose–Einstein condensation of dipolar molecules},
  year      = {2024},
  issn      = {1476-4687},
  month     = Jun,
  number    = {8020},
  pages     = {289--293},
  volume    = {631},
  doi       = {10.1038/s41586-024-07492-z},
  publisher = {Springer Science and Business Media LLC},
}

@Article{Holland2021,
  author    = {Holland, Connor M and Lu, Yukai and Cheuk, Lawrence W},
  journal   = {New Journal of Physics},
  title     = {Synthesizing optical spectra using computer-generated holography techniques},
  year      = {2021},
  issn      = {1367-2630},
  month     = Mar,
  number    = {3},
  pages     = {033028},
  volume    = {23},
  doi       = {10.1088/1367-2630/abe973},
  publisher = {IOP Publishing},
}

@InCollection{Fitch2021,
  author    = {N.J. Fitch and M.R. Tarbutt},
  publisher = {Academic Press},
  title     = {Chapter Three - Laser-cooled molecules},
  year      = {2021},
  editor    = {Louis F. Dimauro and Hélène Perrin and Susanne F. Yelin},
  pages     = {157-262},
  series    = {Advances In Atomic, Molecular, and Optical Physics},
  volume    = {70},
  doi       = {https://doi.org/10.1016/bs.aamop.2021.04.003},
  issn      = {1049-250X},
  url       = {https://www.sciencedirect.com/science/article/pii/S1049250X21000033},
}

@Article{Wright2022,
  author    = {Wright, Sidney C. and Doppelbauer, Maximilian and Hofsäss, Simon and Christian Schewe, H. and Sartakov, Boris and Meijer, Gerard and Truppe, Stefan},
  journal   = {Molecular Physics},
  title     = {Cryogenic buffer gas beams of AlF, CaF, MgF, YbF, Al, Ca, Yb and NO – a comparison},
  year      = {2022},
  issn      = {1362-3028},
  month     = Nov,
  number    = {17-18},
  volume    = {121},
  doi       = {10.1080/00268976.2022.2146541},
  publisher = {Informa UK Limited},
}

@Article{Jarvis2018,
  author    = {Jarvis, K. N. and Devlin, J. A. and Wall, T. E. and Sauer, B. E. and Tarbutt, M. R.},
  journal   = {Physical Review Letters},
  title     = {Blue-Detuned Magneto-Optical Trap},
  year      = {2018},
  issn      = {1079-7114},
  month     = Feb,
  number    = {8},
  pages     = {083201},
  volume    = {120},
  doi       = {10.1103/physrevlett.120.083201},
  publisher = {American Physical Society (APS)},
}

@Article{Bell2009,
  author    = {Bell, Martin T. and P. Softley, Timothy},
  journal   = {Molecular Physics},
  title     = {Ultracold molecules and ultracold chemistry},
  year      = {2009},
  issn      = {1362-3028},
  month     = Jan,
  number    = {2},
  pages     = {99--132},
  volume    = {107},
  doi       = {10.1080/00268970902724955},
  publisher = {Informa UK Limited},
}

@InCollection{Augenbraun2023,
  author    = {Benjamin L. Augenbraun and Loïc Anderegg and Christian Hallas and Zack D. Lasner and Nathaniel B. Vilas and John M. Doyle},
  booktitle = {Advances in Atomic, Molecular, and Optical Physics},
  publisher = {Academic Press},
  title     = {Direct laser cooling of polyatomic molecules},
  year      = {2023},
  editor    = {Louis F. DiMauro and Hélène Perrin and Susanne F. Yelin},
  pages     = {89-182},
  series    = {Advances In Atomic, Molecular, and Optical Physics},
  volume    = {72},
  doi       = {https://doi.org/10.1016/bs.aamop.2023.04.005},
  issn      = {1049-250X},
}

@Misc{Schindewolf2025,
  author    = {Schindewolf, Andreas and Hertkorn, Jens and Stevenson, Ian and Ciardi, Matteo and Gross, Phillip and Wang, Dajun and Karman, Tijs and Quemener, Goulven and Will, Sebastian and Pohl, Thomas and Langen, Tim},
  title     = {From few- to many-body physics: Strongly dipolar molecular Bose-Einstein condensates and quantum fluids},
  year      = {2025},
  copyright = {arXiv.org perpetual, non-exclusive license},
  doi       = {10.48550/ARXIV.2512.14511},
  publisher = {arXiv},
}

@Article{Zeng2026,
  author    = {Zeng, Zixuan and Yang, Shoukang and Deng, Shuhua and Yan, Bo},
  journal   = {Physical Review Letters},
  title     = {Direct Loading of BaF Molecules with a Conveyor-Belt Magneto-optical Trap},
  year      = {2026},
  issn      = {1079-7114},
  month     = Feb,
  number    = {7},
  volume    = {136},
  doi       = {10.1103/pd77-s994},
  publisher = {American Physical Society (APS)},
}

@Article{Lasner2025,
  author    = {Lasner, Zack D. and Frenett, Alexander and Sawaoka, Hiromitsu and Anderegg, Loïc and Augenbraun, Benjamin and Lampson, Hana and Li, Mingda and Lunstad, Annika and Mango, Jack and Nasir, Abdullah and Ono, Tasuku and Sakamoto, Takashi and Doyle, John M.},
  journal   = {Physical Review Letters},
  title     = {Magneto-Optical Trapping of a Heavy Polyatomic Molecule for Precision Measurement},
  year      = {2025},
  issn      = {1079-7114},
  month     = Feb,
  number    = {8},
  pages     = {083401},
  volume    = {134},
  doi       = {10.1103/physrevlett.134.083401},
  publisher = {American Physical Society (APS)},
}

\clearpage

\appendix

\section{Temperature Calibration}
\label{supp:tempcalibration}

\begin{figure}[htb]
        \centering
        \includegraphics[width=0.75\linewidth]{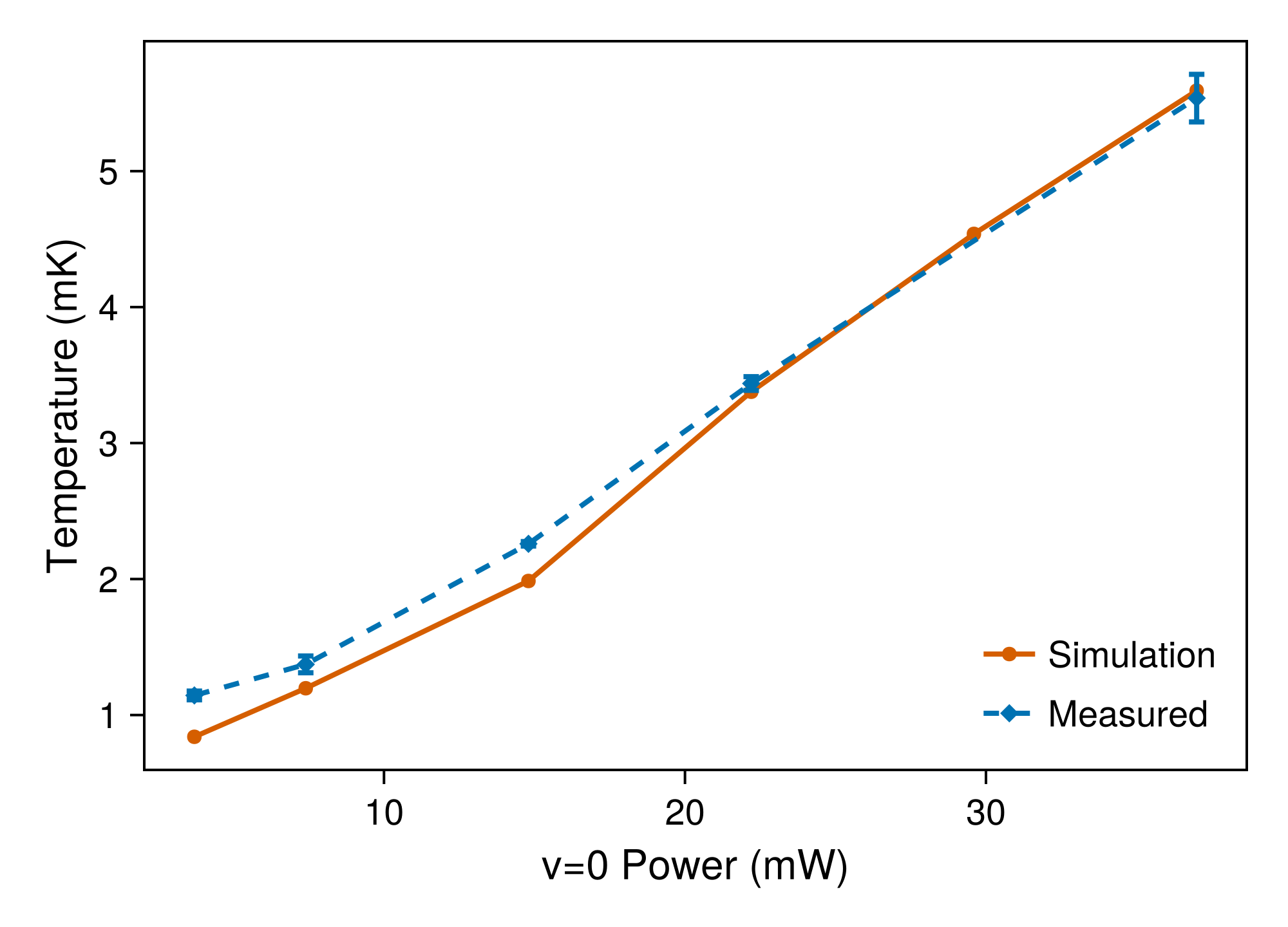}
    \caption{\textbf{Steady-state MOT temperature versus MOT mainline power}. $\Delta = -10~\text{MHz}, ~B_z=-37.5~\text{G/cm}$, $P_{v=1}=43$~mW, $w=$ 8.25~mm.}
        \label{fig:tempcalibration}
\end{figure}

Figure \ref{fig:tempcalibration} shows the experimental (solid line) and simulation (dashed line) temperatures at different MOT beam powers. We see excellent agreement both for the qualitative behavior of the temperature as the intensity is increased and the actual temperature values. For the experimental values, we used a standard time-of-flight measurement to find the temperature of the MOT cloud. To do so, we turn off the MOT coils and lasers and allow the MOT cloud to freely expand, then image the MOT after a variable time $t_{\text{TOF}}$. The imaging pulse is 1 ms long and uses low intensity resonant light as to not cause significant heating of the cloud. The imaged cloud at each $t_{\text{TOF}}$ is then fitted along each of the radial and axial dimensions to a 1D Gaussian to find its size $\sigma$. The temperature along each dimension is found from the slope of $\sigma^2$ vs $t_{\text{TOF}}^2$. The reported temperatures are the geometric mean of the axial and radial temperatures, calculated using $T=T_{\text{radial}}^{2/3}T_{\text{axial}}^{1/3}$. The simulation temperature values were calculated by fitting the velocity norm histogram of the steady-state trapped molecules to a Maxwell-Boltzmann distribution.

\section{MOT Number Calibration}
\label{supp:numbercalibration}
\begin{figure}[htb]
    \centering
    \begin{subfigure}{0.3\textwidth}
        \centering
        \includegraphics[width=\linewidth]{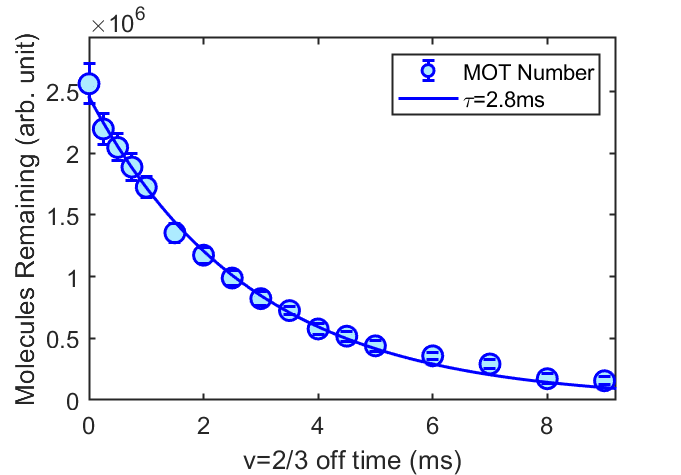}
        \caption{}
        \label{fig:scatteringrate}
    \end{subfigure}
    \begin{subfigure}{0.175\textwidth}
        \includegraphics[width=\linewidth]{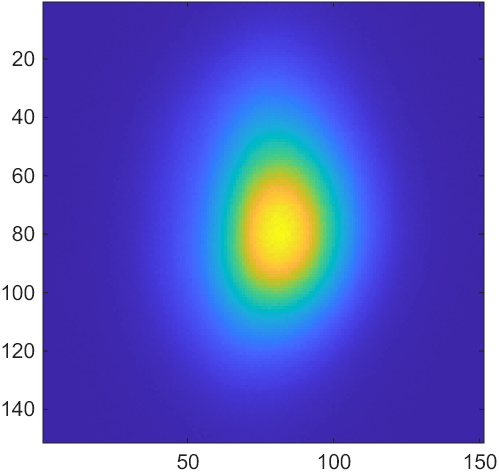}
        \caption{}
        \label{fig:MOTimage}
    \end{subfigure}
    \caption{(a) Number of remaining molecules as a function of time during which the v = 2 and v = 3 repump lasers are turned off. The data shows an exponential decay due to finite photon budget.  (b) Average of 66 fluorescence images of the MOT containing 1.5 million molecules}
    \label{fig:MOTcalibration}
\end{figure}

To calibrate the number of molecules we load into our MOT, we measure the scattering rate and collection efficiency of our system. The scattering rate is measured by turning off the v = 2 and 3 repump lasers and measuring the decay time scale. This reduces the photon budget to 1350. An exponential fit to the decay timescale yields a lifetime of $\tau=2.8$ms, Figure \ref{fig:scatteringrate}. This corresponds to a scattering rate of 480 kHz at the imaging parameters.

The florescence from the MOT is captured onto an EMCCD camera. We calibrated the imaging system to convert camera counts to photons. With this information, and using a 5 ms exposure imaging pulse, we calculate the number of molecules in the MOT. Figure \ref{fig:MOTimage} shows the averaged fluorescence image of the MOT containing 1.5 million molecules.

\section{Other Data}
\label{supp:otherdata}
\begin{figure}[htb]
    \centering
    \begin{subfigure}{0.23\textwidth}
        \centering
        \includegraphics[width=\linewidth]{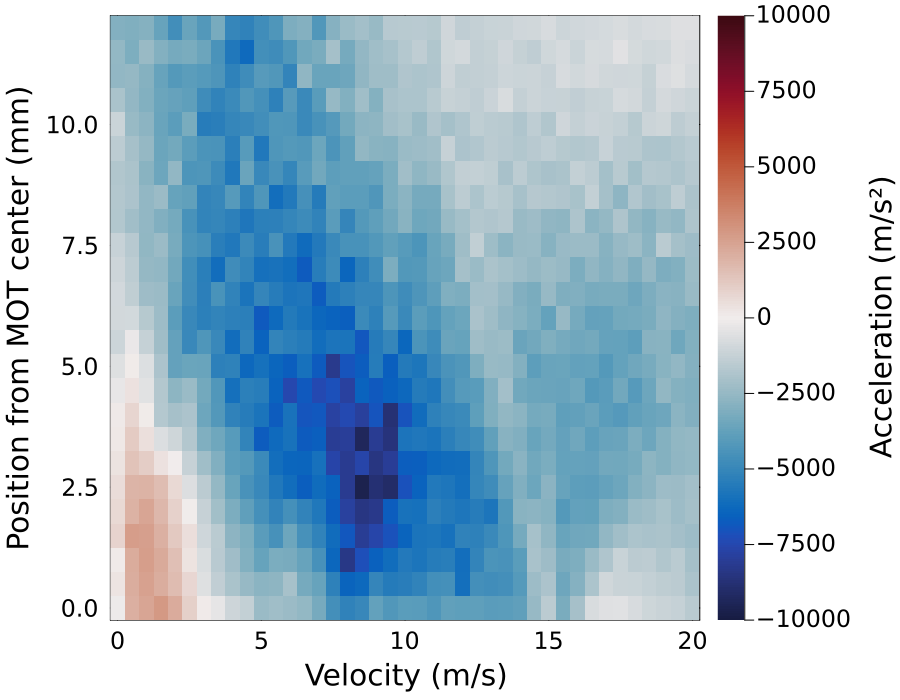}
        \caption{}
        \label{fig:accelerationr0}
    \end{subfigure}
    \begin{subfigure}{0.23\textwidth}
        \includegraphics[width=\linewidth]{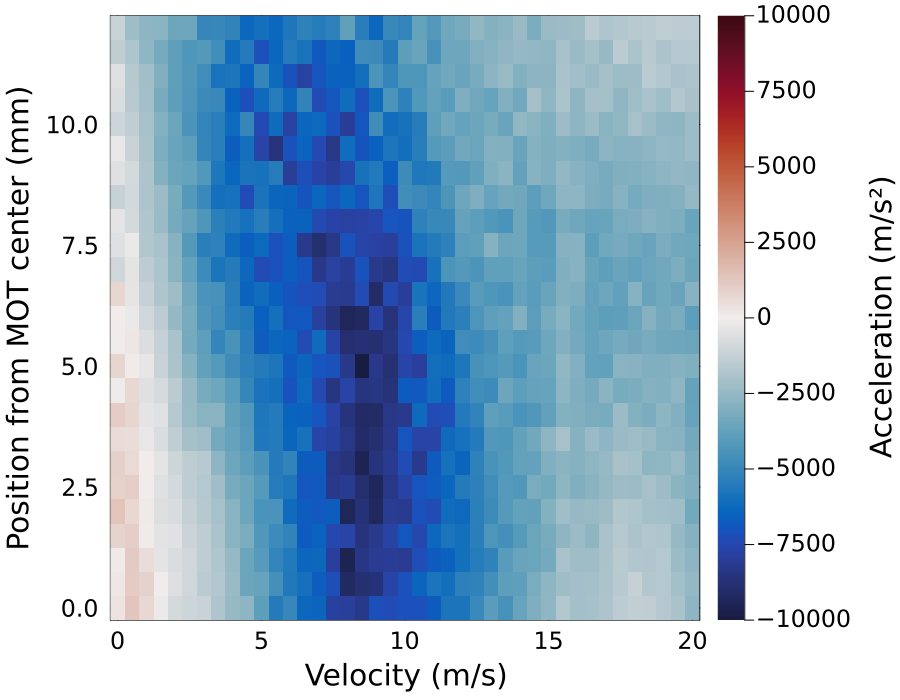}
        \caption{}
        \label{fig:accelerationr4}
    \end{subfigure}
    \begin{subfigure}{0.23\textwidth}
        \includegraphics[width=\linewidth]{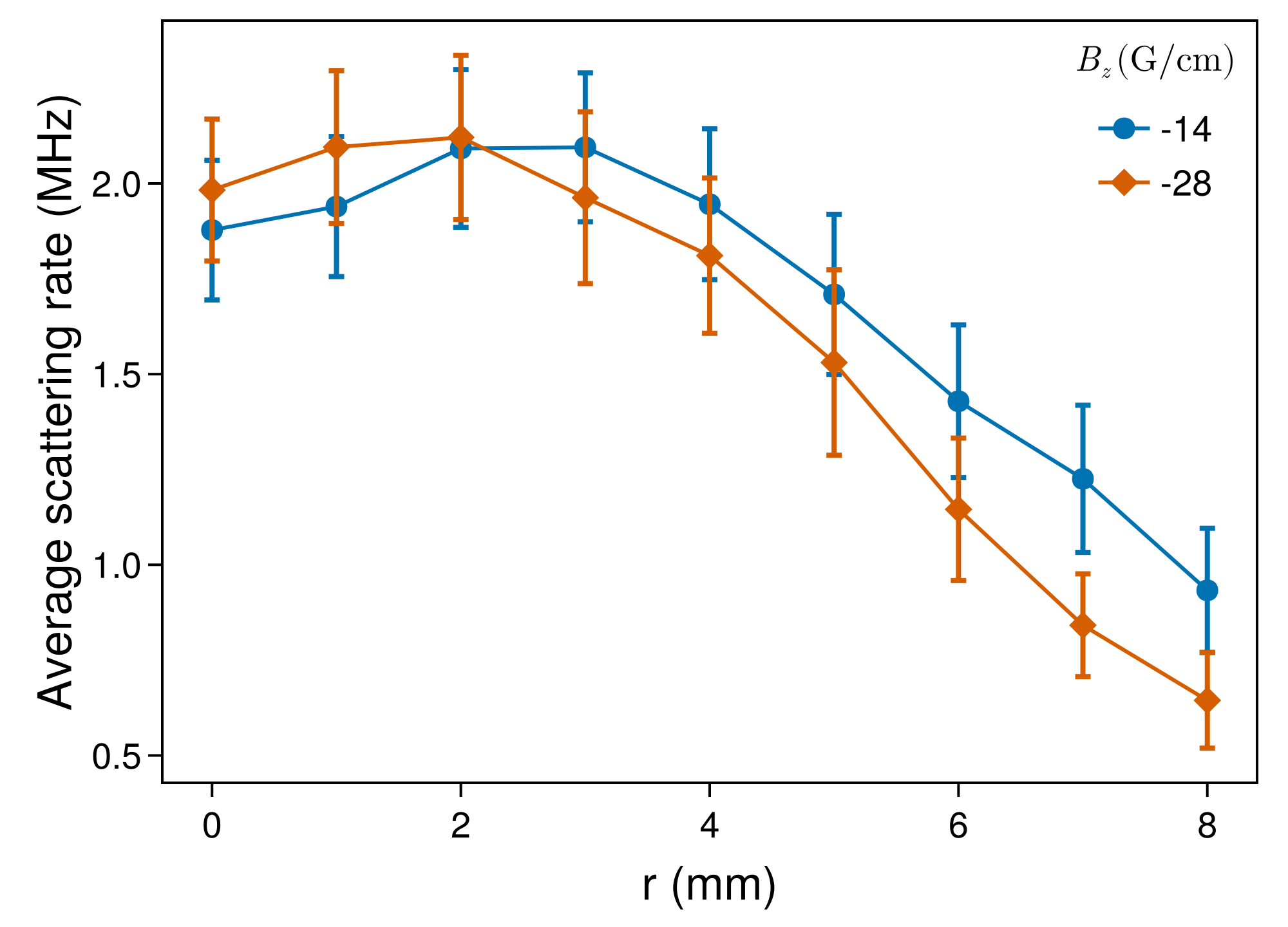}
        \caption{}
        \label{fig:rvsgradientscattering}
    \end{subfigure}
    \begin{subfigure}{0.23\textwidth}
        \includegraphics[width=\linewidth]{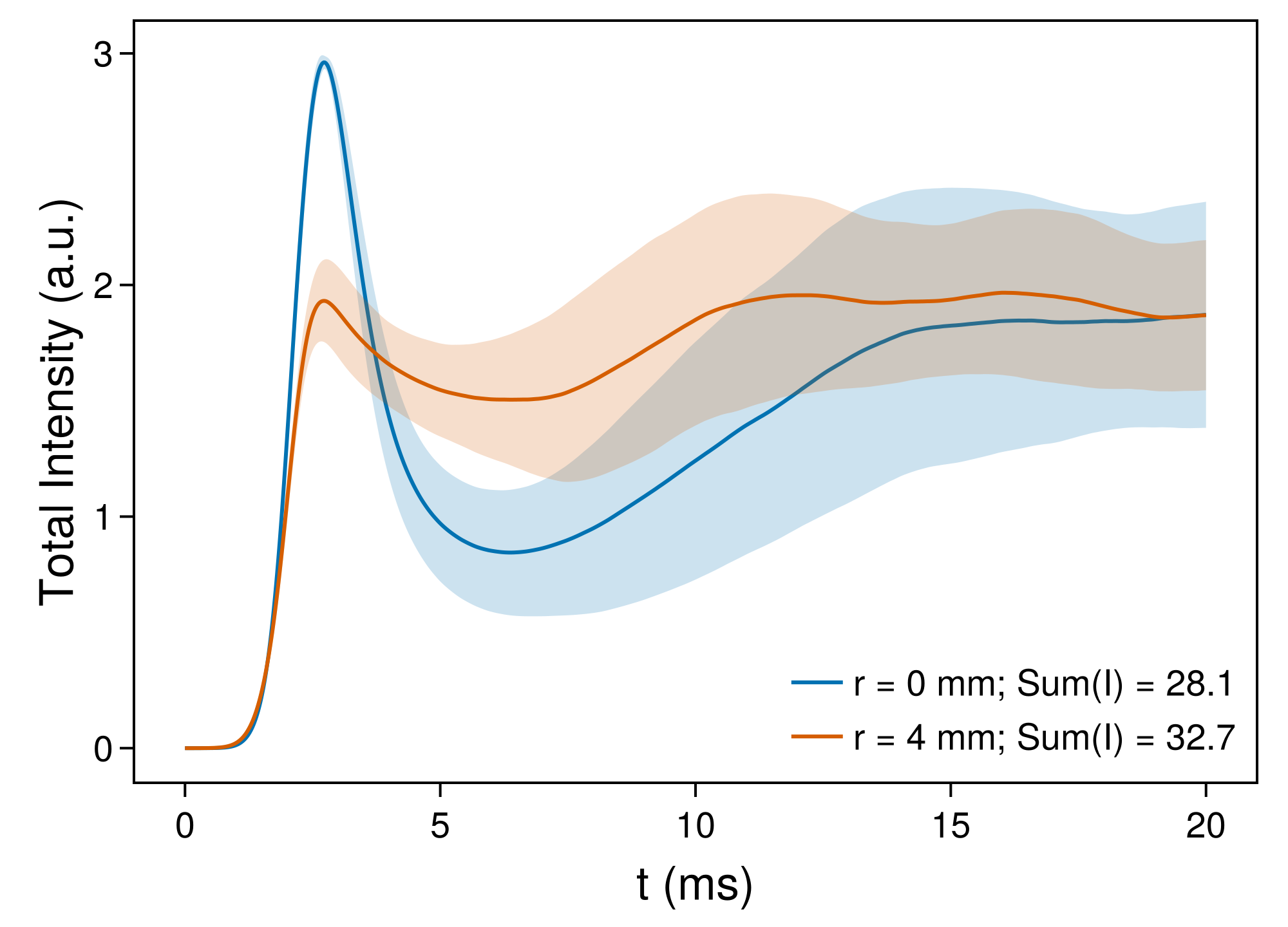}
        \caption{}
        \label{fig:intensityvsr}
    \end{subfigure}
    \caption{\textbf{The increased MOT force, scattering rate, and total intensity at a nonzero radial displacement}. (a) Acceleration as a function of molecule velocity and position in the MOT along $\hat{x'}$ for r=0. Parameters: $B_z=-14~\text{G/cm},~ P_{v=0}=P_{v=1} = 40~\text{mW},~\Delta=-10~\text{MHz},~w=8~\text{mm}$.(b) r=4~mm. (c) Ensemble-average scattering rate for a 12 m/s molecule at $x'=0$ as a function of r for different gradients. (d) The total Gaussian intensity in the trajectory of 12 m/s molecules for r=0 and r=4~mm. Solid lines are the intensity vs time traces averaged over 75 particles. Shaded areas represent standard deviations.}
    \label{fig:rsupportingfig}
\end{figure}

\textbf{\textit{Supporting data for Figure \ref{fig:radiusvsr0}}} Figure \ref{fig:rsupportingfig} serves to explain the increased capture velocity of the MOT for r=4~mm molecules compared to r=0 (on-axis) molecules. Figure \ref{fig:accelerationr0}(\subref{fig:accelerationr4}) shows the acceleration in the MOT as a function of a molecule's velocity and position along $\hat{x'}$ for r=0(4~mm) molecules when $w=8$~mm and $B_z=-14$~G/cm. We see a clear increase in the magnitude of the restoring and cooling forces for the r=4~mm case, in addition to the restoring feature extending over a larger range of positions along x' and the heating feature shrinking. Figure \ref{fig:rvsgradientscattering} shows the scattering rate as a function of r for x'=0, v=12 m/s molecules at $B_z=-14,-28$~G/cm. At both gradient values, we see a slight increase in the scattering rate at a nonzero radial displacement. Additionally, the radial position where the scattering rate is maximized roughly agrees with the capture velocity peaks in Figure \ref{fig:vc_vs_r0} for these two gradient values. Figure \ref{fig:intensityvsr} shows the total Gaussian intensity as a function of time ($I_{tot}(t) = \frac{I_x(y(t),z(t))+I_y(x(t),z(t))+I_z(x(t),y(t))}{I_0}$) experienced by 12 m/s molecules with r=0 and r=4~mm. We see that although the on-axis molecules see a higher maximum intensity due to traveling straight through the center of the MOT, the trajectory taken by the r=4~mm molecules results in a higher sum of the intensity they experience before the reaching equilibrium.

\begin{figure}[htb]
    \centering
    \begin{subfigure}{0.23\textwidth}
        \centering
        \includegraphics[width=\linewidth]{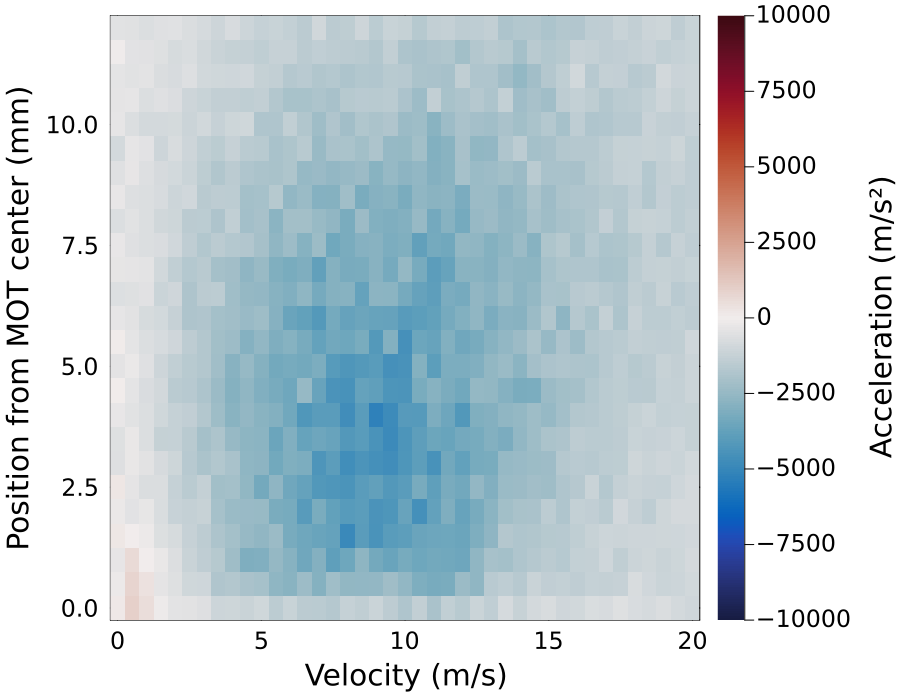}
        \caption{}
        \label{fig:acceleration14G2w}
    \end{subfigure}
    \begin{subfigure}{0.23\textwidth}
        \includegraphics[width=\linewidth]{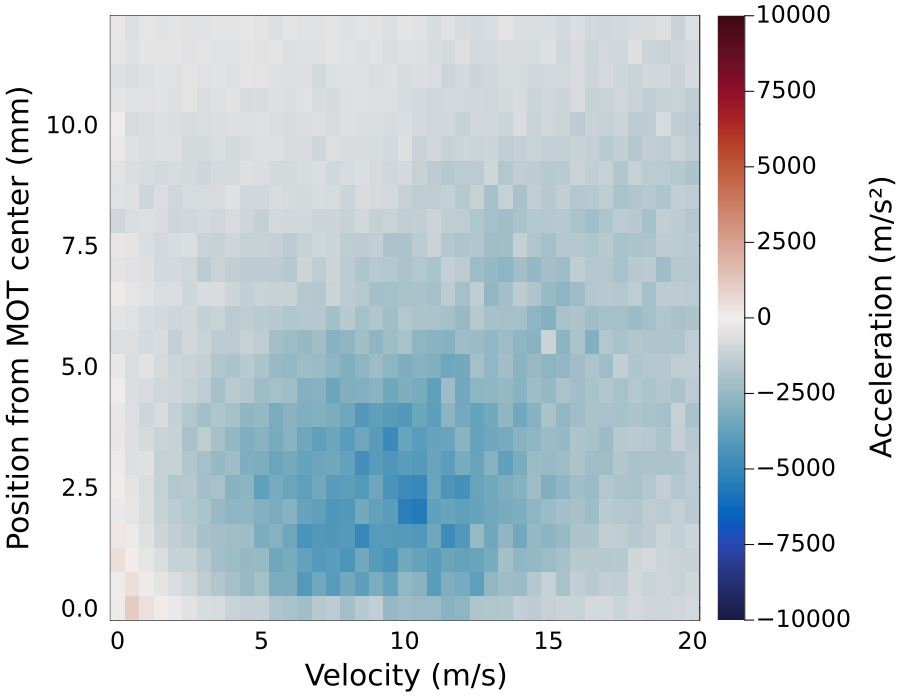}
        \caption{}
        \label{fig:acceleration28G2w}
    \end{subfigure}
    
    \caption{\textbf{Acceleration as a function of molecule velocity and position in the MOT along $\hat{r}$.} 
    Fixed parameters: $w=16~\text{mm}$, $P_{v=0}=P_{v=1}=40~\text{mW}$, $\Delta=-10~\text{MHz}$. (a) $B_z=$ -14~G/cm. (b) $B_z=$ -28~G/cm.}
    \label{fig:largewhighgradientsupport}
\end{figure}

\textbf{\textit{Supporting data for Figures \ref{fig:radiusvsgradient}, \ref{fig:radiusvsgradientvsdetuning}}} Figure \ref{fig:largewhighgradientsupport} can be used to explain the monotonic decrease in the capture velocity of the $2w_0$ beams as the magnetic field gradient increases. By comparing the acceleration maps in Figures \ref{fig:acceleration14G2w} and \subref{fig:acceleration28G2w}, we find that the vanishing MOT force as r increase occurs at a smaller r when $B_z=-28$~G/cm compared to -14~G/cm. This leads to the outer layer of the MOT at a high gradient producing little-to-no force compared to lower gradients, while the inner layers produce roughly the same force, which reduced the amount of power that contributes to the formation of the MOT.

\begin{figure}[htb]
    \centering
    \includegraphics[width=\linewidth]{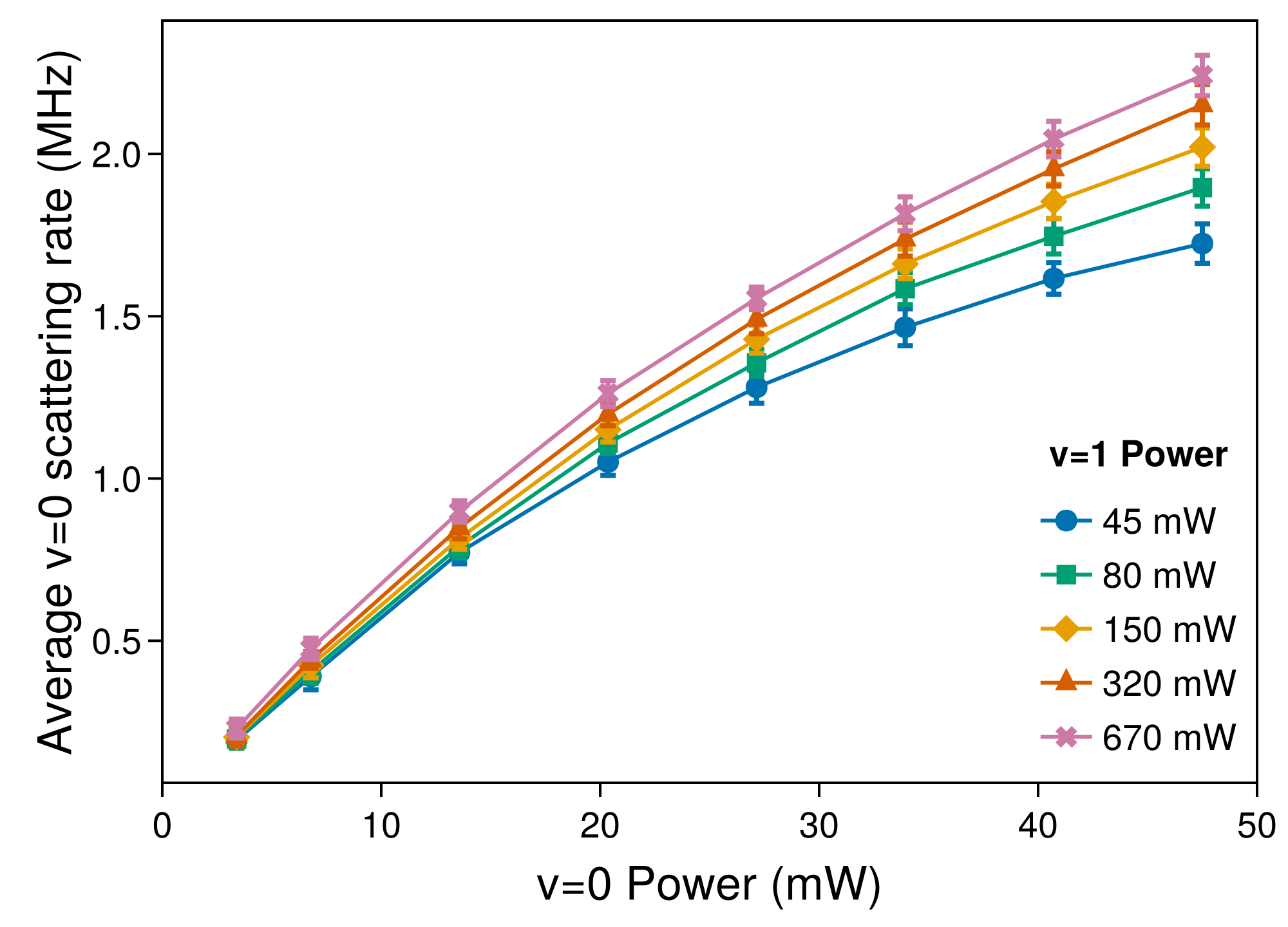}
    \caption{\textbf{Average scattering rate on the X state v=0 branch.} The v=0 scattering rate inside the MOT as a function of the mainline and repump powers at steady state. Scatter points show the average values from an ensemble of 75 particles, and error bars show standard deviations.}
    \label{fig:simulationscatteringrate}
\end{figure}

\textbf{\textit{Supporting data for Figures \ref{fig:accelerationmaps}e and \ref{fig:powerscancalibration}}} Figure \ref{fig:simulationscatteringrate} shows the average scattering rate on the ground state v=0 branch at different v=0 and v=1 powers. For each point in the scan, a stationary ensemble of 75 particles was initialized in the center of the MOT and was allowed to thermalize for 40 ms. The value for the scattering rate is found by counting the events of stochastic collapse to one of the v=0 states. The sum of such counts is then divided by the total time spent inside the MOT. We see a noticeable increase to the v=0 scattering rate as the v=1 power increases, and the gain becomes more prominent at higher v=0 powers.

\begin{figure*}[tbp]
    \centering
    \begin{subfigure}{0.635\textwidth}
        \includegraphics[width=\linewidth]{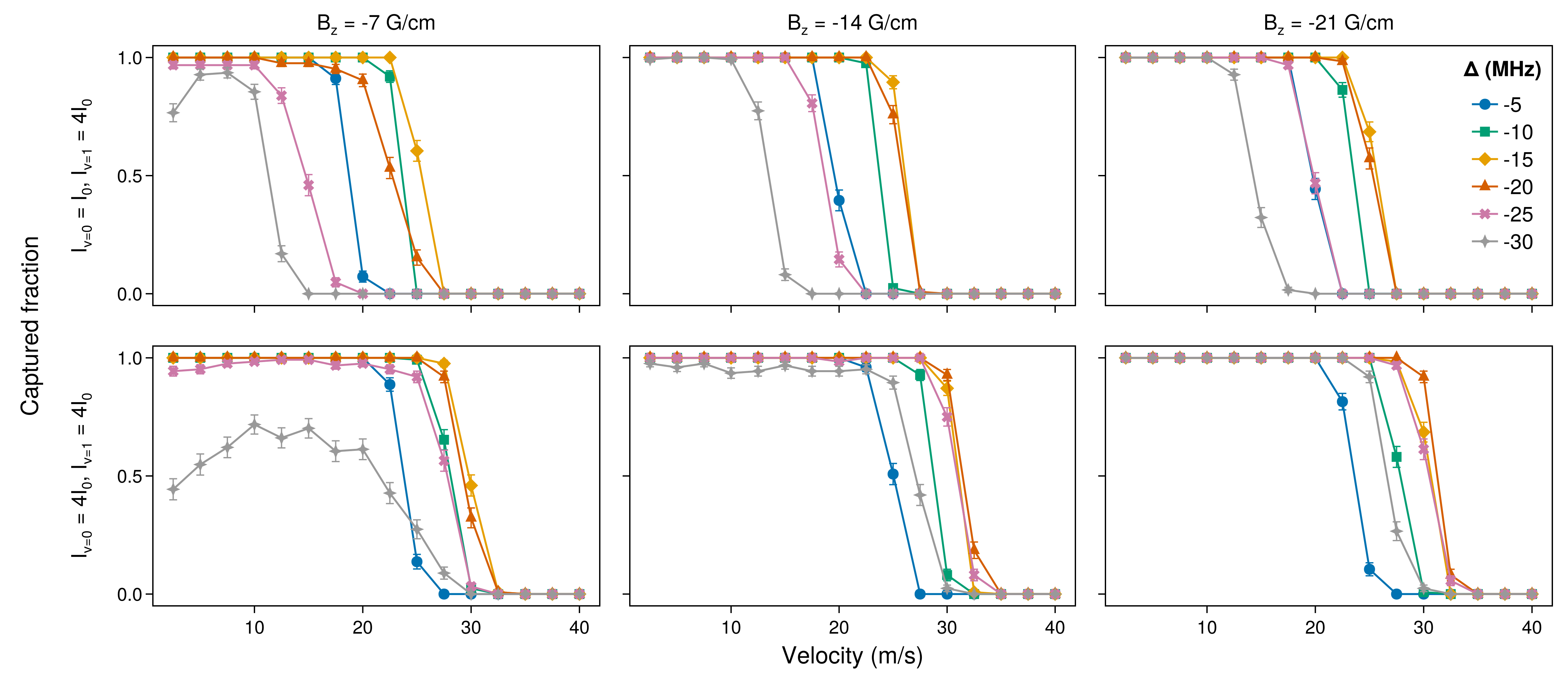}
        \caption{}
        \label{fig:2inchscan}
    \end{subfigure}
    \begin{subfigure}{0.35\textwidth}
        \includegraphics[width=\linewidth]{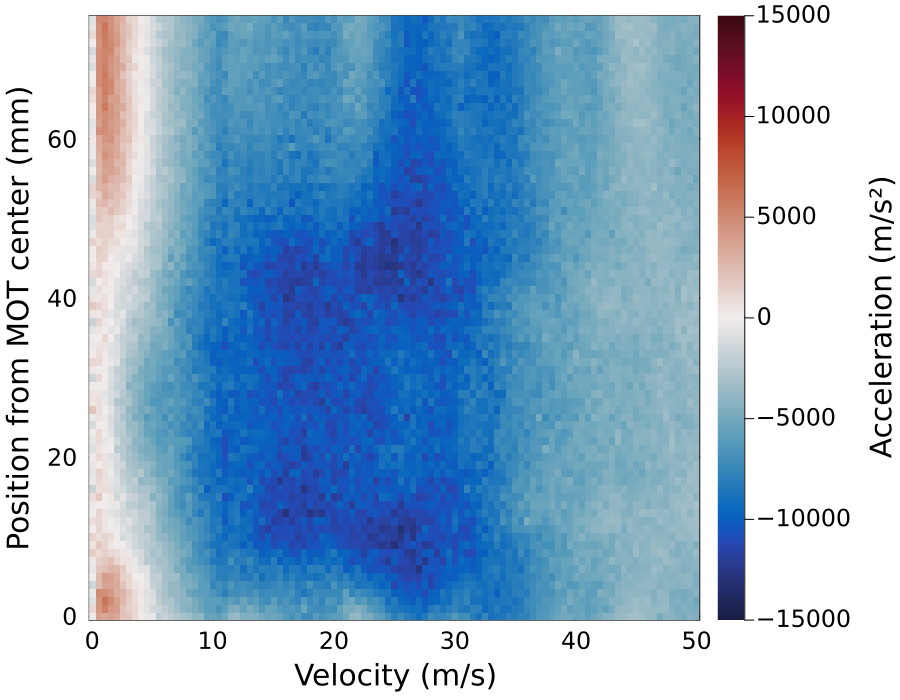}
        \caption{}
        \label{fig:2inchacceleration}
    \end{subfigure}
    \caption{\textbf{Capture velocity in a 2-inch MOT beam system.} (a) Each subplot displays the fraction of molecules trapped at the simulation's end as a function of mainline intensity $I_{v=0}$, magnetic field gradient $B_z$, global detuning $\Delta$, and velocity $v$. Molecules are uniformly launched over a 1 cm diameter molecular beam. Error bars are binomial standard deviations.
    Fixed parameters: $w=22~\text{mm}$, $a=1.15$, and $I_0=40~\text{mW/cm}^2$. (b) Acceleration as a function of molecule velocity and position in the MOT along $\hat{x'}$. An infinite-sized beam is used here. Parameters used: $I_{v=0}=I_{v=1}=160~\text{mW/cm}^2$, $B_z=-14~\text{G/cm}$, and $\Delta=-20~\text{MHz}$.}
    \label{fig:2inchbeams}
\end{figure*}

\begin{figure*}[tbp]
    \centering
    \includegraphics[width=0.75\linewidth]{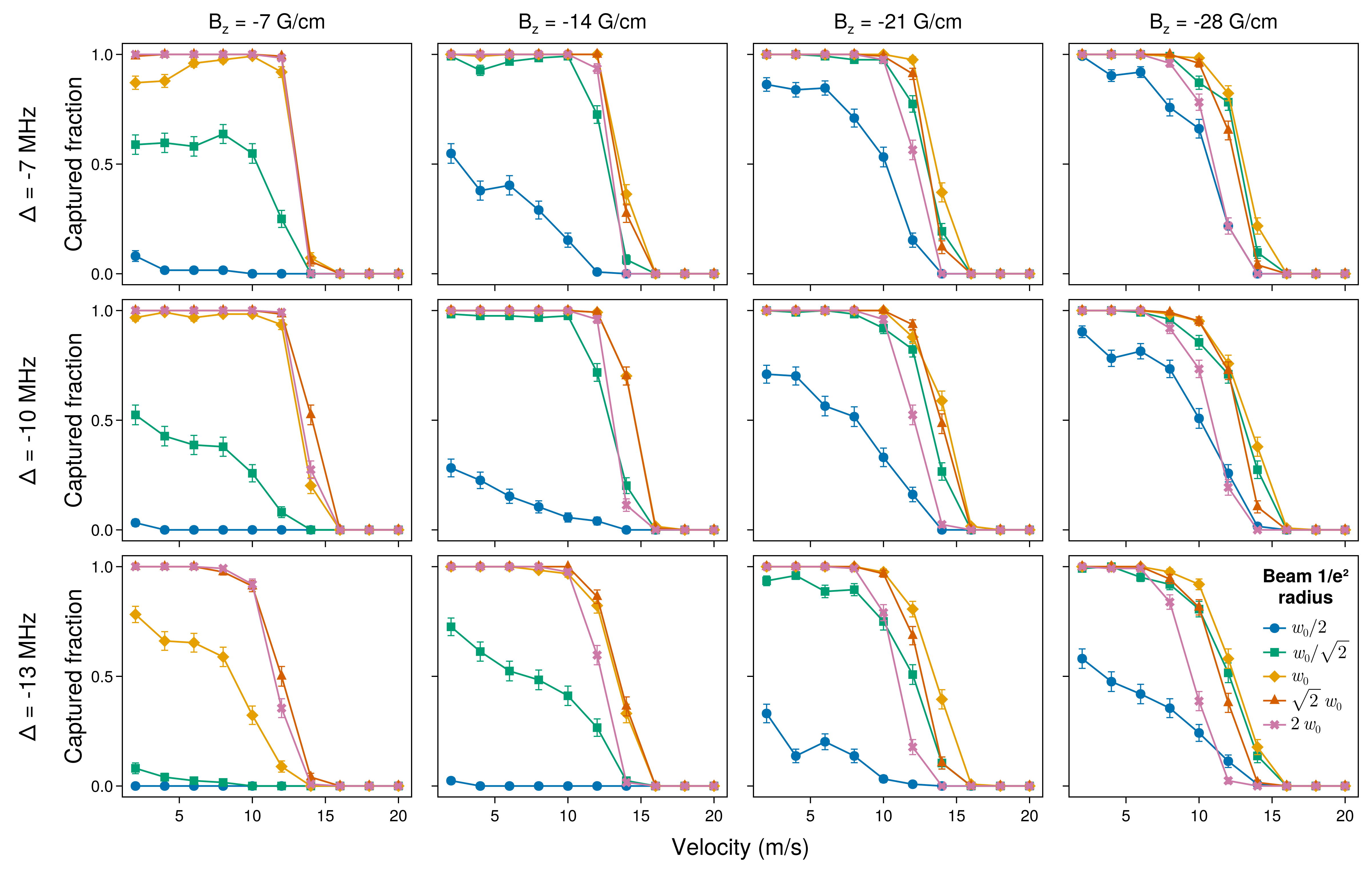}
    \caption{\textbf{Full MOT capture velocity versus beam size at various magnetic field gradients and detunings used for data in Figure \ref{fig:radiusvsgradientvsdetuning}.} Each subplot displays the fraction of molecules trapped at the simulation's end as a function of laser beam size, $w$, magnetic field gradient $B_z$, global detuning $\Delta$, and velocity $v$. Molecules are uniformly launched over a 1 cm diameter molecular beam. Error bars are binomial standard deviations.
    Fixed parameters: $w_0=8~\text{mm}$ and $P_{v=0}=P_{v=1}=40~\text{mW}$.}
    \label{fig:fullradiusvsgradientvsdetuning}
\end{figure*}

\begin{figure*}[tbp]
    \centering
    \includegraphics[width=0.75\linewidth]{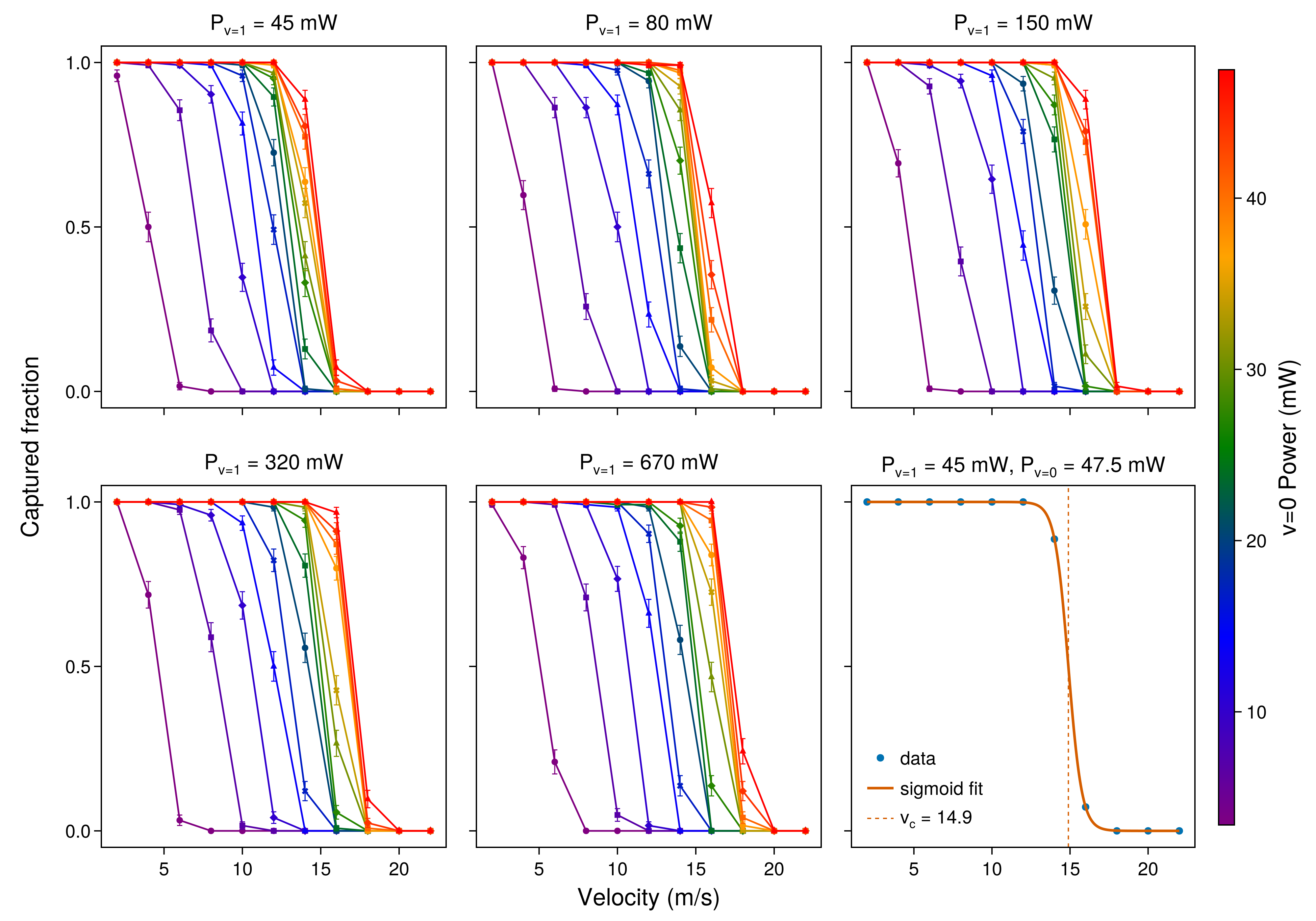}
    \caption{\textbf{Full MOT capture velocity versus mainline and repump power used for data in Figure \ref{fig:simulationpowerscan}.} Each subplot displays the fraction of molecules trapped at the simulation's end as a function of initial velocity and mainline power at different repump powers. Molecules are uniformly launched over a 1 cm diameter molecular beam. Error bars are binomial standard deviations. The bottom right panel shows an example of the sigmoid fit for one of the velocity traces. Fixed parameters: $w=11$~mm, $B_z=-14$~G/cm, $\Delta = -10$ MHz, and $a=1.16$, which corresponds to a 1-inch-diameter iris at this beam size.}
    \label{fig:fullP0vsP1expsimdata}
\end{figure*}

\textbf{\textit{Large capture velocity in 2-inch-diameter MOT beams.}} Figure \ref{fig:2inchscan} shows the capture velocity that can be realized when the MOT beams are scaled to have a radius of $22~\text{mm}$ and an iris factor of $a=1.15$, which corresponds to a 2-inch optical system. We find that at $B_z=-14~\text{G/cm}$, $\Delta=-20~\text{MHz}$, and peak laser intensities $I_{v=0}=I_{v=1}=160~\text{mW/cm}^2$, the MOT can reach a capture velocity of 31.5~m/s. We note that the per-beam power required to reach these intensities at this beam size is 1.2~W, which is higher than available in most experiments. However, this serves to show that given enough laser powers, the capture velocity of molecular MOTs can increase by more than a factor of 2. Depending on the velocity profile of the slowed molecular beam, this increase can in turn lead to another order-of-magnitude boost to the number of molecules trapped in the MOT. Figure \ref{fig:2inchacceleration} shows the acceleration experienced by a molecule at the optimal MOT parameters found in Figure \ref{fig:2inchscan} as a function of position and velocity along $\hat{x'}$. We see a two-lobe region that exhibits strong confining and damping forces. The upper lobe arises from the secondary dual-frequency force created by the $F=1^+$ and $F=0$ states at magnetic field magnitudes $\geq30~$Gauss. In addition, the confining force never completely vanishes between the two lobes, despite the level crossing that occurs at 10 Gauss. We also see that at this detuning, the restoring force regions can extend to velocities higher than 30 m/s. 

\textbf{\textit{Supporting data for Figure \ref{fig:radiusvsgradient}}} Figure \ref{fig:fullradiusvsgradientvsdetuning} shows the full velocity traces used to find the capture velocity values plotted in Figure \ref{fig:radiusvsgradientvsdetuning}.

\textbf{\textit{Supporting data for Figure \ref{fig:simulationpowerscan}}} Figure \ref{fig:fullP0vsP1expsimdata} shows the full velocity traces used to find the capture velocity values plotted in Figure \ref{fig:simulationpowerscan}. The bottom right panel shows an example of the sigmoid fit for one of the velocity traces.

\end{document}